\documentclass[alpha-refs]{wiley-article}

\usepackage{color}
\usepackage{ulem}

\usepackage{siunitx}
\graphicspath{ {./figures/} }
\usepackage{caption}
\usepackage{subcaption}

\papertype{Original Article}
\paperfield{Quarterly Journal of the Royal Meteorological Society }

\title{Improving forecasts of precipitation extremes over Northern and Central Italy using machine learning}

\author[1,3\authfn{1}]{Federico Grazzini}
\author[2]{Joshua Dorrington}
\author[2]{Christian M. Grams}
\author[1]{George C. Craig}
\author[4]{Linus Magnusson}
\author[4]{Frederic Vitart}

\affil[1]{Ludwig-Maximilians-Universität, Meteorologisches Institut, München, Germany}
\affil[2]{Institute of Meteorology and Climate Research (IMKTRO), Department Troposphere Research, Karlsruhe Institute of Technology (KIT), Karlsruhe, Germany}
\affil[3]{ARPAE-SIMC, Regione Emilia-Romagna, Bologna, Italy}
\affil[4]{ECMWF, Shinfield Park, Reading, RG2 9AX, United Kingdom}

\corraddress{Federico Grazzini PhD, Meteorologisches Institut, Ludwig-Maximilians-Universität, München, 80333, Germany}
\corremail{federico.grazzini@lmu.de}

\presentadd[\authfn{1}]{ARPAE-SIMC, Bologna, Emilia-Romagna, 40122, Italy}

\fundinginfo{Ludwig-Maximilians-Universität München; German Research Foundation (DFG), Grant/Award Number: FR4363/1-1; Transregional Collaborative Research Centre, Project T2, Helmholtz Young Investigator Group ‘Sub- Seasonal Predictability: Understanding the Role of Diabatic Outflow’ (SPREADOUT; Grant VH-NG-1243)}

\runningauthor{Grazzini et al. 2023}

\begin{document}

\maketitle

\begin{abstract}
The accurate prediction of intense precipitation events is one of the main objectives of operational weather services. This task is even more relevant nowadays, with the rapid progression of global warming which intensifies these events. Numerical weather prediction models have improved continuously over time, providing uncertainty estimation with dynamical ensembles. However, direct precipitation forecasting is still challenging. Greater availability of machine learning tools paves the way to a hybrid forecasting approach, with the optimal combination of physical models, event statistics, and user-oriented post-processing. Here we describe a specific chain, based on a random forest pipeline, specialised in recognizing favourable synoptic conditions leading to precipitation extremes and subsequently classifying extremes into predefined types. The application focuses on Northern and Central Italy, taken as a testbed region, but is seamlessly extensible to other regions and timescales. The system is called MaLCoX (Machine Learning model predicting Conditions for eXtreme precipitation) and is running daily at the Italian regional weather service of ARPAE Emilia-Romagna. MalCoX has been trained with the ARCIS gridded high-resolution precipitation dataset as the target truth, using the last 20 years of the ECMWF re-forecast dataset as input predictors. We show that, with a long enough training period, the optimal blend of larger-scale information with direct model output improves the probabilistic forecast accuracy of extremes in the medium range. In addition, with specific methods, we provide a useful diagnostic to convey to forecasters the underlying physical storyline which makes a meteorological event extreme. 

\end{abstract}

\section{Introduction}
Italy is among the European nations most exposed to torrential rainfall and flash flooding, due to its geographical conformation and climatological characteristics \citep{ediss28219}. Every year these weather hazards produce huge costs and deadly consequences. The correct prediction of intense meteorological phenomena is one of the main objectives of operational weather services, and this task is even more relevant today with the rapid progression of global warming which amplifies extremes \citep{Seneviratne2021WeatherChange,Tramblay2018FutureMediterranean}. Numerical weather prediction (NWP) models have improved continuously over time, however, providing accurate, quantitative precipitation forecasts remains challenging. Precipitation is an intermittent and complex phenomenon, very often characterized by high spatial variability. Direct predictability of this field is limited compared to the predictability of large-scale circulation patterns in which precipitation events are developing. The broader usage of machine learning in many sectors characterized by high dimensional data, raises the question of whether machine learning can effectively be used for advanced post-processing of meteorological fields, thereby helping to recover the loss of predictability inherent to surface fields characterized by high variability. Research in this direction is rapidly developing with studies showing how machine learning can be applied to advanced postprocessing of precipitation direct model output \citep{Espeholt2022DeepForecasts,Frnda2022ECMWFLearning,deSousaAraujo2022ExtremeBrazil,Whan2018ComparingMethods}, or testing the inclusion of large-scale components such as atmospheric rivers \citep{Chapman2022ProbabilisticLearning}. The link between extreme precipitation events (EPEs) over the Mediterranean area and large-scale atmospheric flow patterns has long been studied and is consolidated in literature \citep{Rudari2005Large-scaleItaly,Grazzini2007PredictabilityAlps,Martius2008Far-upstreamSouth-side,Grazzini2015AtmosphericPackets}. Recently \cite{Mastrantonas2022WhatForecasts} has shown that inferring EPE probability from predefined specific weather patterns outperforms precipitation output in the medium-range, extending the forecasting horizon of the model up to 3 days in many Mediterranean locations. Starting from this perspective, we consider whether machine learning methods can be used operationally to improve extreme precipitation predictions in Italy, combining large-scale dynamical predictors with precipitation direct model output. 
In this work, we describe a random forest (RF) based post-processing chain, called MaLCoX (Machine Learning model predicting Conditions for eXtreme precipitation), specialised in recognizing favourable synoptic and large-scale conditions leading to precipitation extremes and subsequently classifying the categories proposed by \cite{Grazzini2020ExtremeTechniques}. The focus is on Northern and Central Italy, taken as a preliminary testbed region. The choice of predictors is based largely on previous work by the authors \cite{Grazzini2020ExtremeTechniques, Grazzini2021ExtremePrecursors}. In addition, we include non-local predictors: spatial composites of Euro-Atlantic anomaly patterns in the days preceding Italian extreme precipitation events, as described in a recent companion paper \cite{Dorrington2023Domino:Rainfall}. MaLCoX --composed of two modules which detect and classify precipitation extremes respectively -- has been implemented semi-operationally at ARPAE-SIMC using as predictors the available fields the institute is receiving in real-time from the European Centre for Medium-Range Weather Forecasts (ECMWF) dissemination stream. The goal is to set up an innovative `warning bell chain' complementing existing forecasting products, such as direct probabilities and the extreme forecast index \citep{Tsonevsky2015NewConvection.}. This type of hybrid modelling is relatively novel and with promising application in the field of extreme event early warnings, potentially anticipating major events at a time-scale of several days, compared to the current standard of 48h. To our knowledge, this is the first documented machine-learning application for the prediction of extreme precipitation targeted at the medium-range.
The paper is organised as follows: in section 2 we describe the datasets, algorithms, architecture and predictors. In section 3 we show the results of the comparison against direct model output, while in section 4 we illustrate the new MaLCoX forecasting tools applied to a recent case study. Finally, in section 5, we draw our conclusions.

\section{Data and methods}

\begin{figure}[!ht]
\centering
\includegraphics[width=15cm]{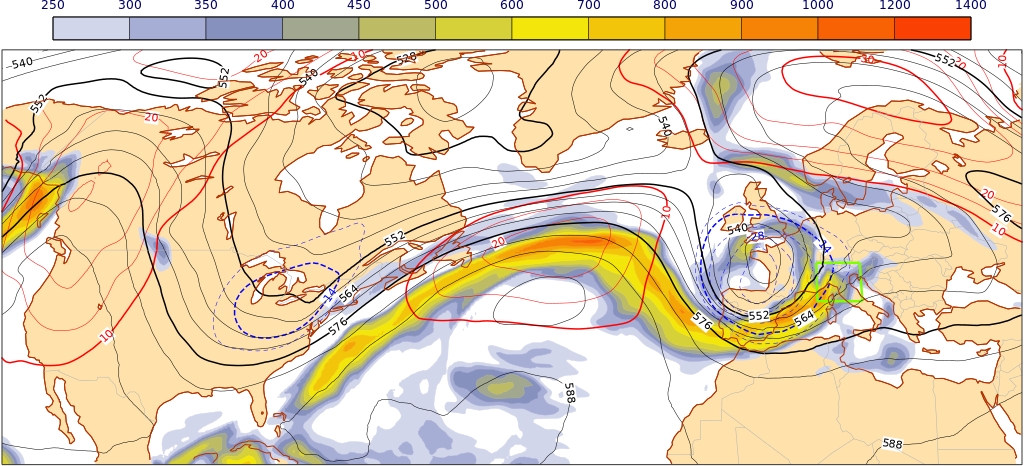}
\caption{Synoptic situation 2020-10-02 12:00 UTC associated with storm Alex. An EPE event occurred inside the test region depicted with the green box between the 2nd and 3rd of October 2020. The map displays the geopotential height at 500 hPa which describes the middle-level flow (solid lines) and the associated magnitude of the instantaneous vertically integrated water vapour flow shaded according to the scale above in $kg s^{-1} m^{-1}$. Blue dashed isolines and red solid lines show the corresponding 500 hPa anomaly for the seasonal climatological average which is then projected onto the non-dimensional Z500 non-local index which in this flow configuration is 2.4.}
\label{fig:example_case_Alex}
\end{figure}

To introduce the geographic area and its relation with synoptic predictors used in the model, in Fig.\ref{fig:example_case_Alex} we show the synoptic situation associated with storm Alex, chosen as an exemplary case. We are not going to discuss the forecast of this event but use it to illustrate the concept of non-local predictors referred to later. A trough is deepening over western Europe, at the leading edge of an incoming Rossby wave packet. At the same time, strong integrated water vapour transport (IVT) is connecting the upstream trough over the U.S. east coast and the deepening trough over Western Europe, forming an atmospheric river on the northern side of the Atlantic ridge. This is a typical and recurrent synoptic situation associated with the strongest EPEs, as discussed in the work of \cite{Sioni2023RevisitingItaly} where the detailed evolution of the two most severe EPEs ever recorded over Northern Italy is discussed. On the 2nd and 3rd of October 2020 a large EPE was recorded in the area of interest, inside the green box in Fig.\ref{fig:example_case_Alex}, with significant damages and flooding in the western Alps, in Italy and France, as described in \cite{Magnusson2021WindstormEurope}.

\subsection{Event definition and target variable}
Our definition of EPE consists of an extreme precipitation event occurring within the area indicated by the green rectangle in Fig.\ref{fig:example_case_Alex}. The condition of extreme precipitation yes/no, is obtained through thresholding based on the observational ARCIS dataset. ARCIS is a high-resolution (5x5 km) gridded precipitation dataset obtained by the spatialization of a high-density surface observation network of 11 Italian regions plus several stations of adjacent Alpine regions, described in \cite{Pavan2019High19612015}. Gridded data covers Northern and Central Italy at daily resolution from the first of January 1961 up to real-time. Each day since then has been labelled with EPE yes if the aggregated daily precipitation of one or more of the 94 warning areas which is subdivided Northern and Central Italy (see \citep{Grazzini2020TheItaly} for details) exceeds the 99th percentile of its wet days climatology of the recent climate period (1991-2020) and the sum of the area above the 99th percentile is greater than 1000$km^2$. We focus on medium to large extremes, filtering out very localised downpours. The obtained tabular time series of dates with EPE yes/no is our ground truth or target variable. In the recent period (1991-2022), we have observed 782 EPE days, 24.4 $\pm$ 6.9 EPEs each year, with a seasonal distribution ranging from about $\sim$5\% in winter (DJF), spring (MAM) and summer (JJA) and $\sim$12\% in autumn (SON). 
An equivalent time series of precipitation forecast output, used as a benchmark, is obtained by applying the same thresholding rules to the direct model output daily precipitation, using the 24h short-term precipitation forecast from ERA5 reanalyses as base climatology for computing the 99 percentile thresholds. This results in lower thresholds, in absolute terms, for the direct model outputs to raise an EPE yes.

\subsection{Machine learning algorithm description}
\begin{figure}[!ht]
\centering
\includegraphics[width=12cm]{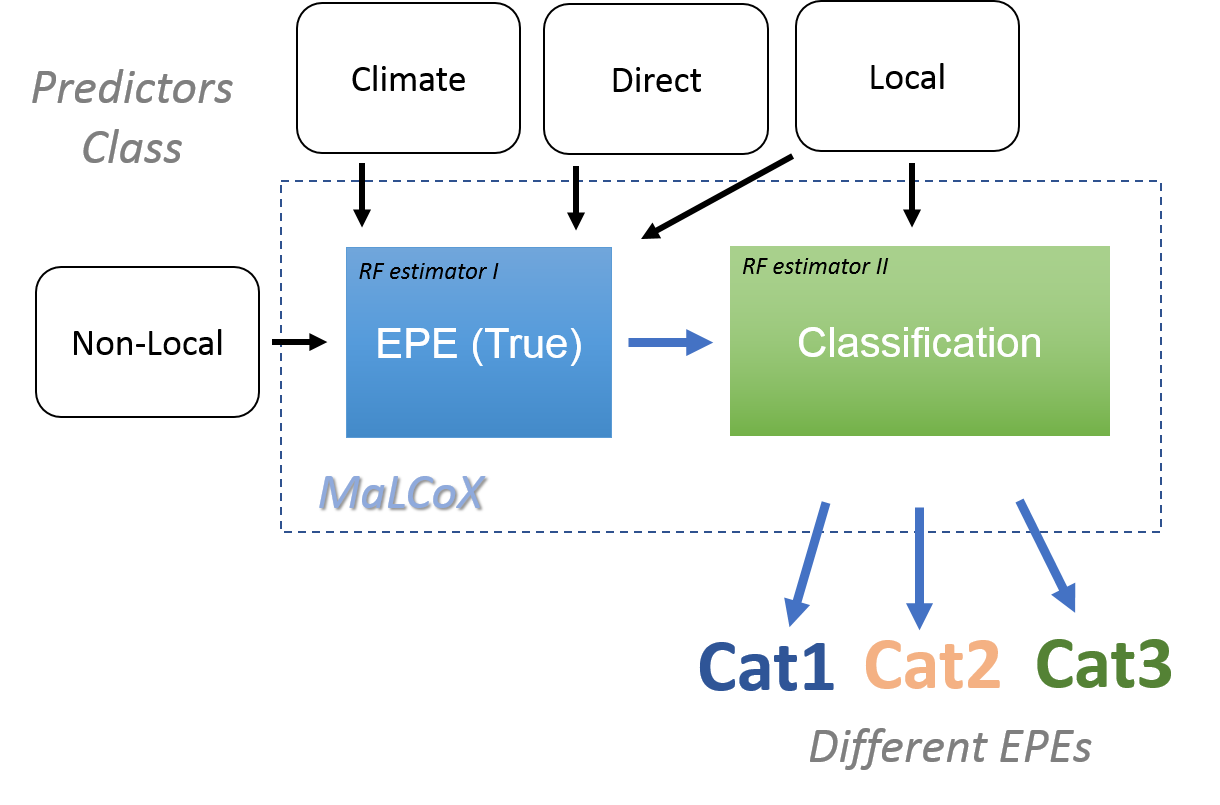}
\caption{Schematic of MaLCoX architecture}
\label{fig:Malcox_schema}
\end{figure}

MaLCox consists of a pipeline of Python modules containing machine learning models and estimators from the  Scikit-Learn library \cite{Pedregosa2011Scikit-learn:Python}. At the core of MaLCoX's algorithms lies the random forest method \citep{Breiman2001RandomForests}. Random forest is an estimator that fits many decision tree classifiers on various randomly drawn sub-samples of the dataset instances and predictors. It uses averaging on the trees to improve the predictive accuracy and control over-fitting. Prediction, which can be categorical, probabilistic or continuous regression, is made by aggregating the individual predictions of the ensemble of decision trees. Random forests are already proven in the context of severe weather detection \citep{Hill2020ForecastingForests} and are straightforward to implement and optimise. The architecture of MaLCoX can be divided into two major blocks, both using RF estimators, schematized in Fig.\ref{fig:Malcox_schema}. A first module (EPE module) is designed to predict the probability of an EPE and eventually the total volume of rain (Vol) and the EPE area extension (EPEarea). A second block (Classification module) classifies EPE into categories defined in \cite{Grazzini2020ExtremeTechniques}. Besides different predictors, the two modules differentiate themselves by their different levels of tuning. While the task of EPE classification is relatively simple (we use the default hyperparameters) since it works on a homogeneous set of dates containing only EPE days, the task of the first module (EPE yes/no) is more challenging and requires some specific hyperparameter customization accounting for EPE rarity. RF produces individual trees from a bootstrap sample of the training data. In learning extremely imbalanced data, as our dataset is, there is a significant probability that a bootstrap sample contains few or even none of the minority class, resulting in an overall poor performance in predicting the positive event (EPE yes). To alleviate this problem, in the EPE module, we set the option of having balanced subsamples which automatically adjust weights inversely proportional to class frequencies in the bootstrap sample for every tree grown. The other problem to face was avoiding over-fitting to improve generalization. Instead of setting manually parameters such as min$\_$samples$\_$leaf and max$\_$depth, we used the cost complexity pruning to control the size of a tree to prevent overfitting. This pruning technique is parameterized by the cost complexity parameter (ccp$\_$alpha) with values greater than zero progressively increasing the number of nodes pruned. We find the optimal value by testing different values recursively (via GridSearch) using cross-validation on the training dataset. For our application, the choice of ccp$\_$alpha=0.001 is the best compromise which maximises the validation scores. While the classification module is independent of lead time, the EPE yes/no module has a different RF model for each forecast step, with a different training dataset and the same hyperparameters. We tested also having the predictors and hyperparameters change with lead time, trying a recursive feature elimination with cross-validation but the results were similar to presenting the same predictors to the model for each lead time and letting the model decide how to use them. We opted then for this latter choice since it is preferable to have the same number of predictors at all lead times facilitating interpretability.
The EPE (yes/no) module includes also two additional RF regressor models used to compute the expected rain volume and EPEarea, fitted with the same predictors of the classification models.
Finally, we complemented the software modules, with the Shapley additive explanations library \citep{Lundbergetal.2017} specifically designed to help explain the outputs of machine learning models. It allows us to effectively visualise the contribution of each feature for a particular prediction. The prototype of MaLCoX has been put in a pre-operational and testing phase, running daily on ARPAE servers since September 2022.

\subsection{Training and test dataset}
On average there are only 24 EPE days for each year, so we need a sufficiently long dataset for the training period to build meaningful statistics. Another requirement, since we are dealing with forecasts of events occurring in different years, is that only small changes in model performances occur over time. These requirements restricted the choice to the ECMWF re-forecast, which is a coarse horizontal resolution (18km), for the investigated period, compared with high-resolution 9km (HRES) operational runs but more stable in terms of model changes. This system is composed of an 11-member ensemble running biweekly for 46 days, with the latest IFS cycle, on the same initial date (day and month) for the previous 20 years \citep{Vitart2019}. We take the available 20-year re-forecast sets produced in 2021, 2020, 2019 to fill gaps introduced by the fact the system is not running daily, accumulating statistics over different model cycles. In case of duplicate dates, we keep only the most recent forecast.
Another restriction is represented by the availability of the IVT fields. This fundamental predictor which only became available in June 2018, from model cycle 45r1 of the ECMWF IFS system output. To reduce the already large data transfer, we train our model only on the control (CTRL) member of the ensemble re-forecast. We also tested training based on the reanalysis but as we want to account for the mean forecast error of the predictors, which is lead time-dependent,  we opted for training on the re-forecast. After removing duplicate re-forecast dates, the training dataset consists of $\sim$5282 days with the number of EPEs slightly changing with forecast steps (357 $\pm$ 12). Forecast steps are spanning from +24h (D1) to +240h (D10). 
The test dataset is assembled on the same set of predictors extracted from the ECMWF HRES operational daily runs from July 2018 (the first available cycle with IVT as output) until 2022. All the verification dates already included in the training dataset, about 10\% of the original available HRES dates, are being excluded from the test dataset to avoid any overlap. At the end, the number of days present in the test dataset is 898 of which 58 $\pm$ 4 EPEs depending on the forecast lead time.

\subsection{Non-local, local, direct and climate predictors}
\begin{figure}[!ht]
\centering
\includegraphics[width=16cm]{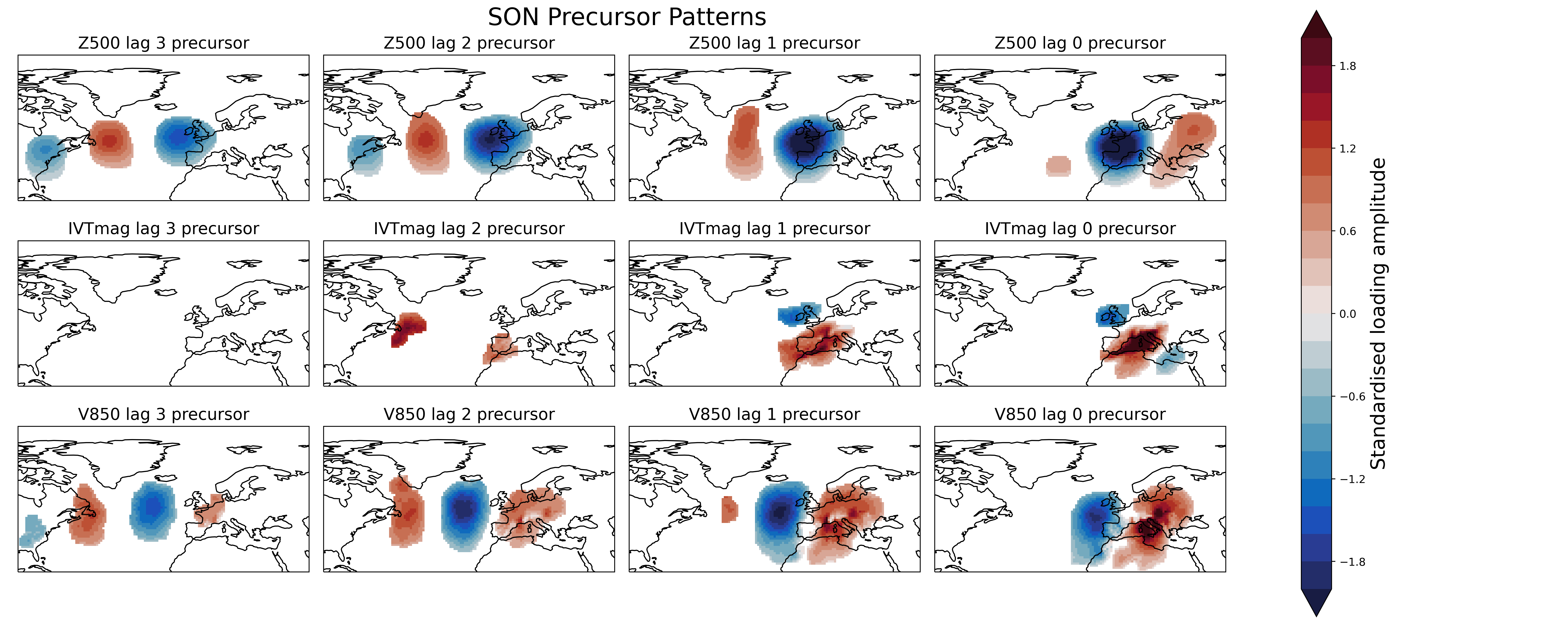}
\caption{Standardised composites of non-local precursors (Z500/top row, IVT magnitude/middle row, V850/bottom row) from 3 days before (lag 3) up to the day (lag 0) of EPEs occurring in SON season}
\label{fig:Precursors}
\end{figure}

We define four sets of predictors: non-local, local, direct and climate predictors. Table \ref{tab_predictors} shows the full list of predictors subdivided according to their class and usage in the two modules of the MaLCoX model. The majority of predictors are used in the first step, in the EPE module, used to detect EPE True/False. The remaining five predictors are used to classify the type of event in three main EPE categories: frontal or orographic uplift of moist statically stable flow (Cat1), stronger frontal uplift of a  neutrally moister/warmer stable flow with embedded convection (Cat2), thermally forced deep convective ascent (Cat3), as proposed in \cite{Grazzini2020ExtremeTechniques}. For the training, all the predictors are obtained from the control member of the re-forecast, at 6-hour intervals up to 15 days and aggregated at daily resolution (last 5 days not used at the moment). In the test period and real-time application the predictors are taken instead from the HRES ECMWF forecast every 6h up to day 10 and aggregated at daily resolution, benefiting from higher resolution. This difference between training (low res) and testing (high res), which makes the comparison unfavourable for MaLCoX, will disappear in the current (after June 2023) IFS model cycles, when the ensemble prediction forecast (ENS) and HRES will have the same resolution and re-forecast resolution will also be upgraded accordingly. 
Let us introduce the first class of predictors: the non-local predictors. This class of predictors is based on the prior systematic identification of precursor patterns, as lagged-anomaly-composites of large-scale variables prior EPEs, defined in ERA5 data, following the approach of \citep{Dorrington2023Domino:Rainfall} and using the associated `Domino' Python package. Patterns are masked based on the statistical significance of anomalies, amplitude and spatial extent of anomalies, and then standardized. Finally, time-evolving indices are computed from the spatial patterns for selected variables (Z500, V850 and IVT magnitude) as the projection (scalar product) of the daily deseasonalised field anomaly onto the precursors patterns. In practical terms, the non-local indices summarize the spatial similarity of current synoptic conditions to the typical precursor patterns associated with EPEs n days before (with n from 0 up to 5 days before the event) over the Euro-Atlantic area. Precursor patterns are computed on a seasonal basis accounting for the change of large-scale dynamics during the year. In Fig.\ref{fig:Precursors} we show an example of precursors patterns for autumn (SON) EPEs. 

The wave pattern associated with EPEs is very evident in the three respective atmospheric variables at day 0 (the same day as the EPE), and it remains coherent while shifting west in the days before the event, up to day -3 in Z500 and V850 and up to day -2 in IVT (Fig.\ref{fig:Precursors}). Coherency and significance depend on the season, increasing in SON and DJF and decreasing in MAM and JJA (not shown). In the current version of the system, we use non-local precursors up to day -2, to have significant amplitudes for all variables. This means that for each valid date, we have up to three non-local indices for each variable, leading to nine independent non-local indices. For example, the Z500 indexes are computed projecting the instantaneous forecast anomaly validating on the composite two days before (lead2), one day (lead1) and the day of the event (lead0). So the non-local predictors of each day are not only a function of fields valid at D0 but also reflect the forecast for the days before; they are non-local both in space and time. Qualitatively, it is possible to see in Fig.\ref{fig:example_case_Alex} how the anomalies of the wave associated with storm Alex, and high IVT values, match the composites of Fig.\ref{fig:Precursors} respectively, with a resulting index for Z500=2.4, V850=3 and IVTmag=3.9 at lag0; very high values corresponding to 2.4 to almost 4 standard deviations compared to the mean values observed for autumn EPEs.

\begin{table}[!ht]
\caption{Table showing the list of predictors. Predictors are subdivided according to their type and usage in the two-step blocks of MalCoX model}
\begin{threeparttable}
\resizebox{\textwidth}{!}{\begin{tabular}{lllll}
\headrow
\thead{Variable} & \thead{Description} & \thead{Units} & \thead{RF model} & \thead{Class}\\
fcst\_IVTmag & IVT normalized anomaly index lead times from 0 to -2 days &  & EPE(y/n) & Non-Local\\
fcst\_Z500 & Gepot. normalized anomaly index at 500 hPa lead times from 0 to -2 days &  & EPE(y/n) & Non-Local\\
fcst\_V850 & 850hPa meridional wind normalized anomaly index at 850 hPa lead times from 0 to -2 days & & EPE(y/n) & Non-Local\\
IVTe & Daily mean of zonal component of integrated water vapour transport & $kg s^{-1} m^{-1}$ & EPE(y/n) & Local\\
IVTn & Daily mean of meridional component of integrated water vapour transport & $kg s^{-1} m^{-1}$ & EPE(y/n)& Local\\
TCWV & Daily mean of total column water vapour & $kg m^{-2}$ & EPE(y/n)& Local\\
Mslp & Daily mean of mean sea level pressure & $hPa$ & EPE(y/n)& Local\\
Volf & Daily volume of rain over the target domain & $m^{3}$ & EPE(y/n)& Direct\\
Juld & Day of the year (Julian day) &  & EPE(y/n)& Climate\\
$\Theta_{e850}$ & Daily mean of equivalent potential temperature at 850hPa & $K$ & Classification & Local\\
$\Delta\Theta_{e500-850}$ & Daily minimum of delta $\Theta{e(500 – 850)}$hPa & $K$ & Classification & Local\\
$\Theta{pv2}$ & Daily mean of $\Theta$ on dynamical tropopause (pv2) & $K$ & Classification & Local\\
Taudmax & Daily maximum of convective adjustment time scale & $h$ & Classification & Local\\
CAPEdmax & Daily maximum of CAPE & $J kg^{-1}$ & Classification & Local\\
\hline  
\end{tabular}}
\end{threeparttable}
\label{tab_predictors}
\end{table}

Local predictors are a set of thermodynamic and dynamic variables, describing the circulation at the local scale, averaged temporally at the daily resolution, and spatially over the green box of Fig.\ref{fig:example_case_Alex}. The choice of variables has been made through a combination of established variables described in a previous work of the authors \cite{Grazzini2020ExtremeTechniques} plus the addition of $\Theta$ on pv2 isosurface, as a tracer of upper-level wave activity. The choice of this variable compared with pv on $\Theta$ isosurfaces instead, as proposed in \cite{Grazzini2021ExtremePrecursors}, is mostly motivated for practical reasons due to the availability of this field in the operational dissemination already in place at ARPAE. We introduced also a direct model predictor, the daily volume of rain forecast (Volf) over the entire target area, as a feature to convey the explicit prediction of the model. An attempt without direct model output showed worse performance, especially in the short-term forecast. Finally, we introduced the climate predictors class, which in the current configuration is composed only by the day of the year. This variable provides direct information on observed EPE frequency which shows a marked seasonal cycle (see Fig.2 of \cite{Grazzini2020ExtremeTechniques}). In addition to this simple information, this variable allows MalCoX to modulate the importance of other predictors according to the time of the year. We tested the introduction of other slow varying predictors in the climatological class, accounting for example for the observed warming trend like sea surface temperature (or its deseasonalized anomaly) averaged at the scale of the Mediterranean basin. We found that Mediterranean averaged SST variables did not add predictive power to the model and therefore we discarded it. A possible explanation is that SST variability occurs at longer time scales compared to synoptic disturbances. According to our predictor's correlation matrix (not shown here), there is no significant correlation (r = 0.03) between EPE area or rain volume (Vol) and SSTs averaged over MedSea. This doesn’t mean that SSTs are not influential in EPE genesis, they are indeed important for heat and momentum fluxes. Simply there there is no covariance because even if SSTs are high, or anomalous, they will stay high also when the phenomena is not occurring (especially in summer). Probably latent heat flux (not tested yet) would be a better indicator, with variability on the same temporal scales of EPEs. In addition, water vapour sources from remote areas are sometimes more important for EPE development, as confirmed by other recent studies \citep{Duffourg2011OriginFrance,Khodayar2022WhatMediterranean,Khodayar2021OverviewHyMeX}, so that the direct Mediterranean sea contribution for some EPEs is less important, like in winter.

\section{Model evaluation}
In this section, we discuss the model performance over a large sample of cases. As a main score to evaluate the performance of MaLCoX, against the direct model output of the ECMWF HRES precipitation taken as a benchmark, we used the average precision score (AP) available from the scikit-learn 1.2.2 library. Receiver Operator Characteristic (ROC) curves are commonly used to present results for binary classification in machine learning. However, when dealing with highly skewed and imbalanced datasets, like in our application, the AP, which is the area under the precision-recall curve, gives a more informative picture, as discussed in \cite{Davis2006TheCurves}. Precision (P) is a metric that quantifies the number of true positive predictions of minority class (EPE = yes) divided by all positive (true plus false), while recall (R), also known as sensitivity, is the fraction of true positive divided by true positives plus false negatives. AP is calculated as follows: 
\begin{equation*}
AP = \sum_{n} (R_{n}-R_{n-1})P_{n}
\end{equation*}
where $P_{n}$ and are $R_{n}$ the precision and recall at the nth threshold.

In addition to AP, we use the Brier score to assess the improvement of the inclusion of the non-local predictors. Brier score or Brier score loss is a negatively oriented score, the smaller the better, which measures the mean squared difference between the predicted probability and the actual outcome. 

\subsection{Feature importance}
\begin{figure}[!ht]
\centering
\includegraphics[width=10cm]{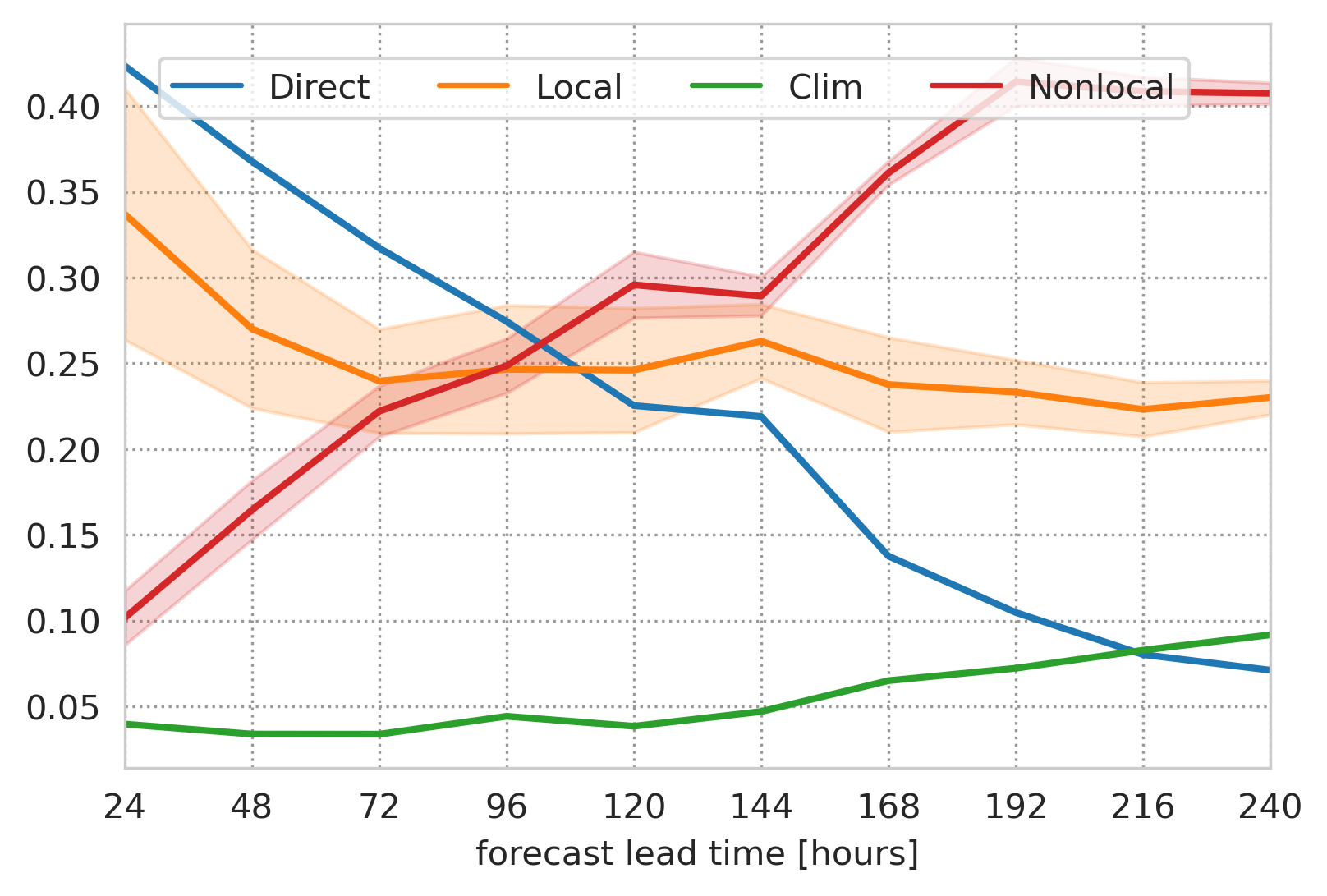}
\caption{RF feature importance evolution with forecast lead time [hours], aggregated by class of predictors according with table 1}
\label{fig:Features}
\end{figure}

Before focusing on the scores, it is useful to discuss the contribution of each feature class to the model. One of the main reasons for conceiving MaLCoX as a hybrid model is that we can study the relative importance of direct versus large-scale predictors for the correct prediction of EPEs at different forecast ranges. This can be done using the feature importance of the RF estimator. As might be expected, in the first three days of the forecast the direct model output of precipitation (Volf) is the dominant predictor. At short ranges, direct rainfall prediction from state-of-the-art physical models is very well correlated with observed precipitation, especially if temporally and spatially aggregated. As we enter the medium-range, precipitation errors grow rapidly and other synoptic variables become more relevant in discriminating days with EPE yes/no. Namely, IVTn is the second most important variable after Volf until D+3, while from D+4 onward the IVTmag (lead0) surpassed IVTn in importance (not shown). To synthesize the relative contribution of the different features, we aggregated the predictors according to their type, indicated in the last column of \ref{tab_predictors}, and show how relative importance changes with forecast lead time. This is shown in Fig.\ref{fig:Features}, where we observe a crossing point between the feature importance of the aggregated predictors by type around D+4. From this point onward, the overall effect of non-local predictors (representing the larger-scales) becomes predominant over the other types. Direct model prediction rapidly decreases in importance, reducing to almost climatological value by D+9, while local predictors importance remains almost constant, mostly supported by the IVTn contribution. The constant increase of climate predictors from D+5 onward is also notable. 

\subsection{Performance in the train and test dataset}
\begin{figure}[!ht]
\centering
\includegraphics[width=10cm]{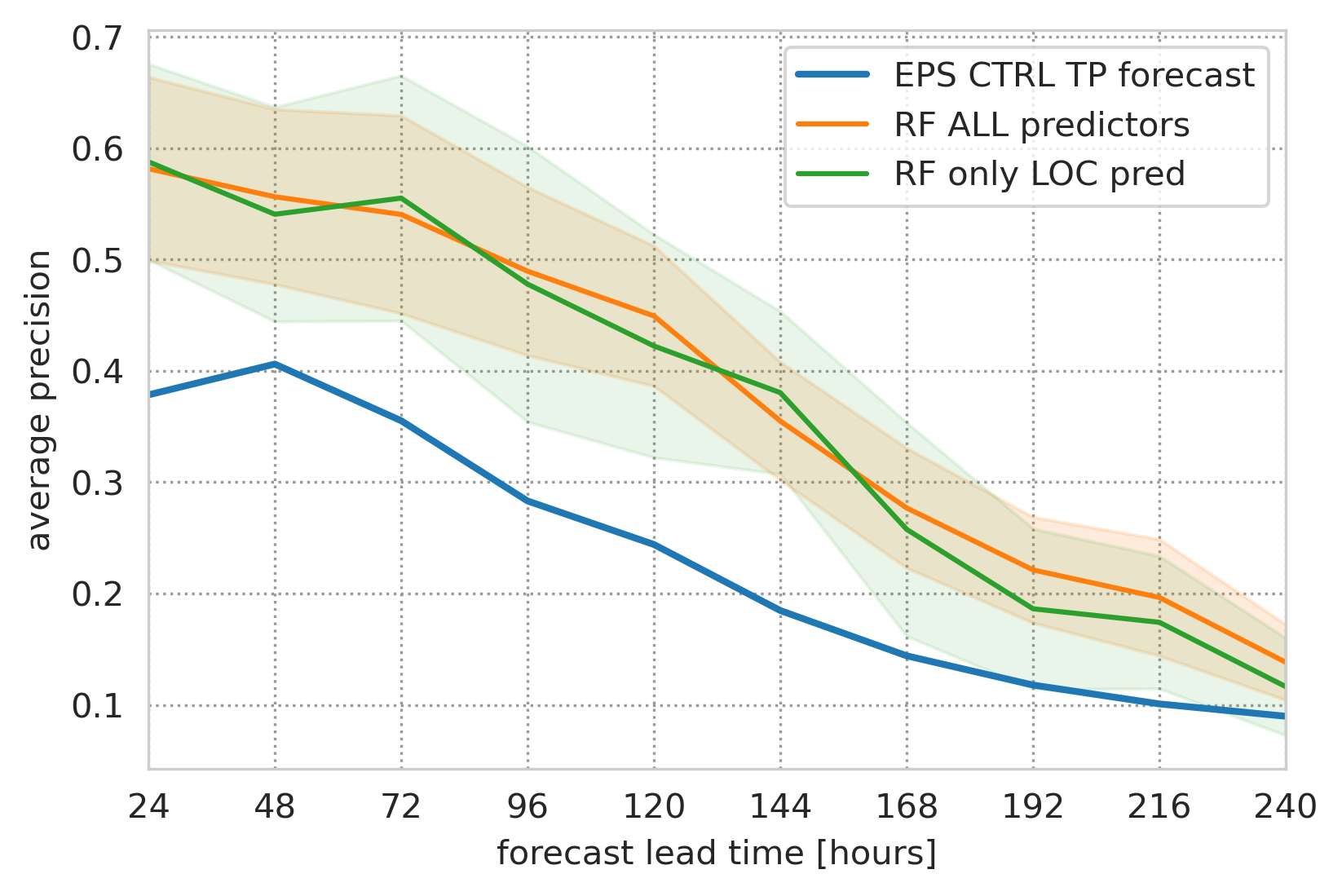}
\caption{Average precision score computed on the training dataset (on average 5282 days, 357 with EPE). Scores averaged over 5 cross-validation folds}
\label{fig:Training_score}
\end{figure}

\begin{figure}[!ht]
\centering
\includegraphics[width=10cm]{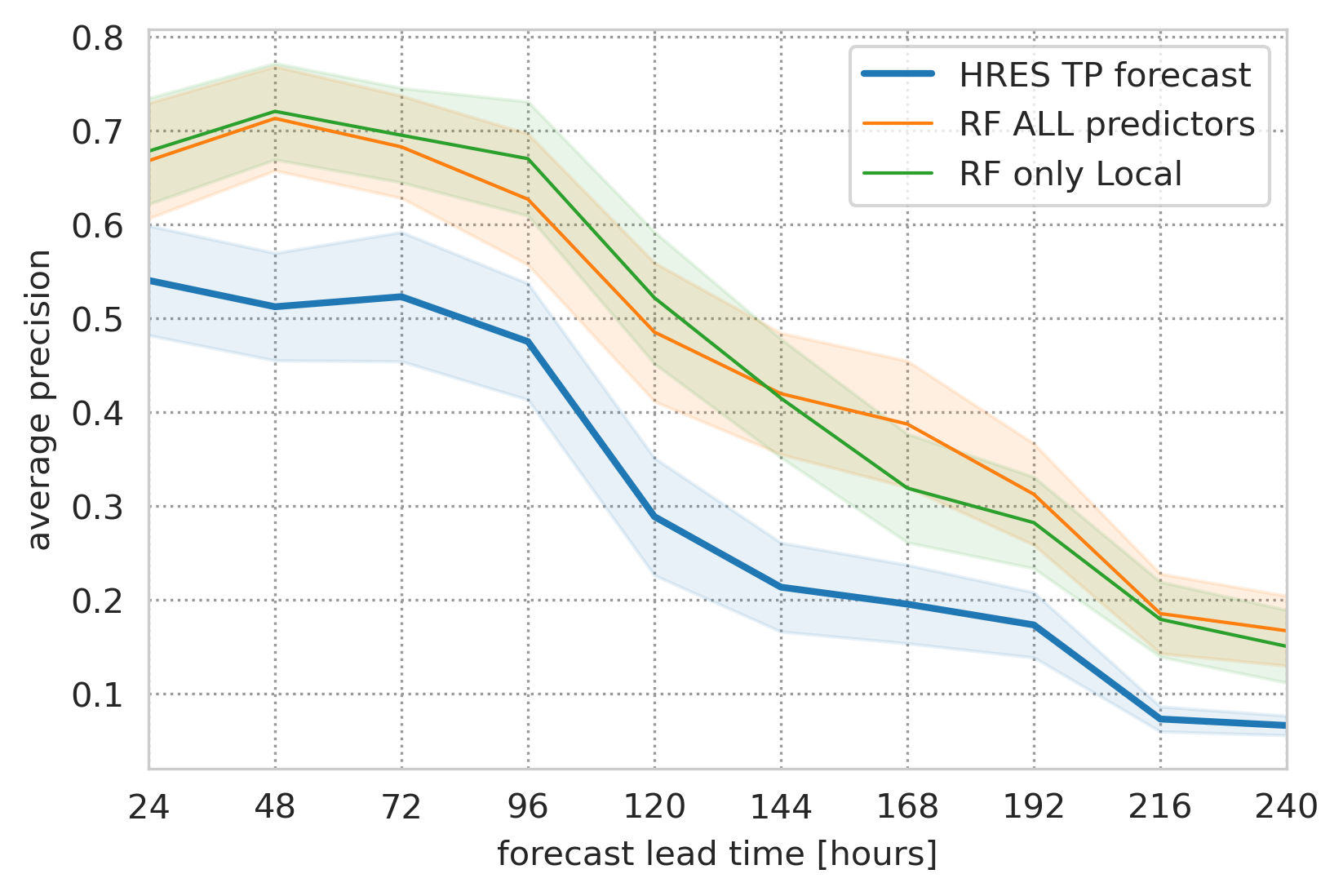}
\caption{Average precision score computed on the test dataset (on average 898 days, 58 with EPE)}
\label{fig:Test_score}
\end{figure}

Once we defined the metrics, we tested two MalCoX versions using different predictors. RF ALL contains all predictors, while RF LOC excludes the non-local predictors. The difference between the models is informative as to the importance of non-local predictors. The performance is first assessed on the training dataset with a cross-validation procedure (CV), in which the training set is split into five smaller folds used to compute the validation metrics after being trained on the remaining folds.
The result of the probabilistic prediction of the two MaLCoX versions for the train and test are also compared respectively against the categorical prediction obtained from the control forecast (CTRL), shown in Fig.\ref{fig:Training_score}, and later the HRES forecast in Fig.\ref{fig:Test_score}. As shown in Fig.\ref{fig:Training_score} MaLCoX performance evaluated in the training dataset is always much better than CTRL with the largest gain in the medium-range between +96 and +144 lead time. Both versions, ALL and LOC, are showing a similar level of skill up to 7 days (forecast lead time +168). After, RF ALL tends to perform marginally better hinting at a positive role of the non-local features for the prediction of EPEs at longer ranges, although there is still a large overlap of confidence intervals which could not allow us to draw firm conclusions.


We apply the same metrics to the test dataset, not seen during the training. The test dataset contains a smaller sample size but is still relevant for the significance of the statistics with 898 days on average and 57 EPEs. To estimate the mean and standard deviation of the scores we applied a 100-time resampling procedure with replacement. From Fig.\ref{fig:Test_score} we see that the HRES precision score is higher than the CTRL score due to higher resolution and accuracy in predicting EPEs. Better skill of HRES precipitation has a positive influence also on MalCoX skill which in general is higher compared with results obtained in the training dataset. RF models, even if trained on the CTRL, continue to show higher precision throughout the forecast period compared to the HRES categorical precipitation forecast. This advantage is greatest in the medium-range, where the skill gain of using MaLCoX instead of HRES precipitation is equivalent to about three days; the skill of MaLCoX (ALL) at D7 forecast is almost the same as HRES direct model output at D+4. Secondly, we see a similar behaviour of the two model configurations observed in the training dataset. RF All shows an even larger advantage, although not significant, compared to RF LOC at forecast lead time +144. It should be noted however that despite AP being an appropriate score there is a disparity in comparing a probabilistic forecast with a categorical benchmark. A more fair comparison should be done against probabilistic prediction obtained from the full ECMWF ensemble over a long time window, planned for a more comprehensive verification in a future study.

\begin{figure}[!ht]
\centering
\includegraphics[width=10cm]{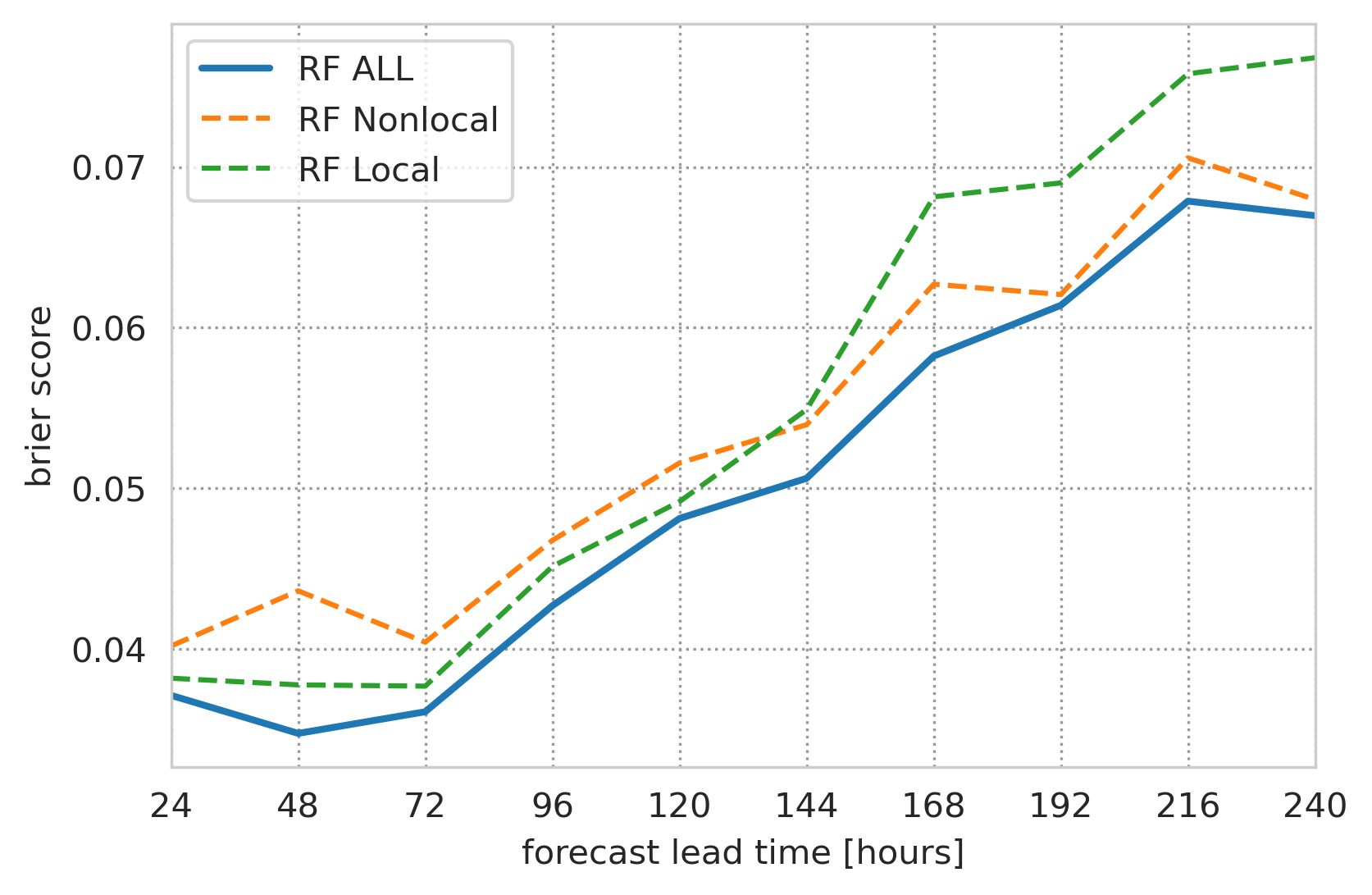}
\caption{Brier score of different EPE models containing different sets of predictors, ALL, only Local (green curve), only Nonlocal (orange curve)}
\label{fig:Brier_score}
\end{figure}

To investigate further the relative importance of the non-local vs local predictors we consider also different skill metrics in the test dataset, among those the Brier score. This skill metric proved particularly effective in highlighting the relative role of different types of predictors that contribute to keeping the performance of MalCoX consistently higher than the precipitation model output. In Fig.\ref{fig:Brier_score} we compare the Brier score, which is negatively oriented, of RF ALL predictors against RF LOC and a further version, denominated RF Nonlocal, which shows only the contributions on non-local predictors. It is interesting to see that the brier score of RF ALL, up to lead time +144, is very similar to the brier score of RF Local (green curve), while after, in the late medium-late range, the level of skill is the same of RF Nonlocal (orange curve). This behaviour seems to reinforce the hypothesis, coming from the feature importance analysis, that non-local predictors are taking over local predictors from D6 onwards.

\begin{figure}[!ht]
\centering
\includegraphics[width=10cm]{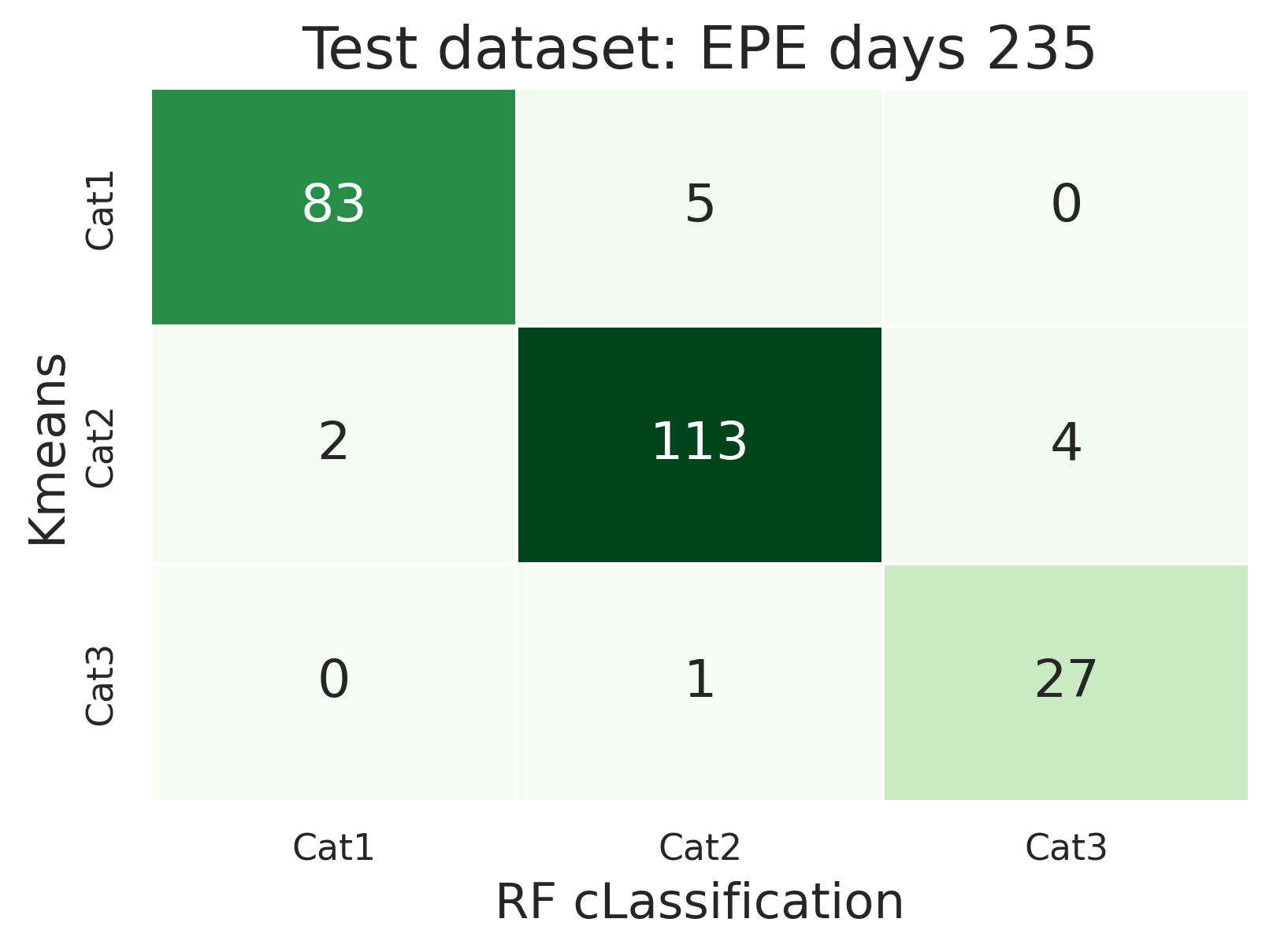}
\caption{Confusion matrix showing the accuracy of prediction of the EPE classification module over the test dataset containing 235 EPE}
\label{fig:Confusion_matrix}
\end{figure}

Finally, we show a measure of the accuracy of the EPE classification model in assigning EPEs (in the forecast) at one of the defined three categories Cat1, Cat2 and Cat3. Note that, differently from the RF models used in the EPE yes/no block, the RF EPE classifier is trained on the analysis fields (1991-2022) using only days where EPEs were observed and a perfect prog approach. Assuming perfect predictors, the skill of the RF category classifier is very good in reproducing the categories obtained by Kmeans clustering, our ground truth, based on the same predictors (see \cite{Grazzini2020ExtremeTechniques} for a description of the clustering). The confusion matrix in Fig.\ref{fig:Confusion_matrix}, shows that about 87\% of the predictions in Cat3 are correct, 95\% for Cat2 and 98\% for Cat1. The test dataset is obtained from a random sampling of 30\% of days in 1991-2022, excluded by the training period.

\section{Case Study: 15 December 2022}
\begin{figure}[!ht]
\centering
\includegraphics[width=10cm]{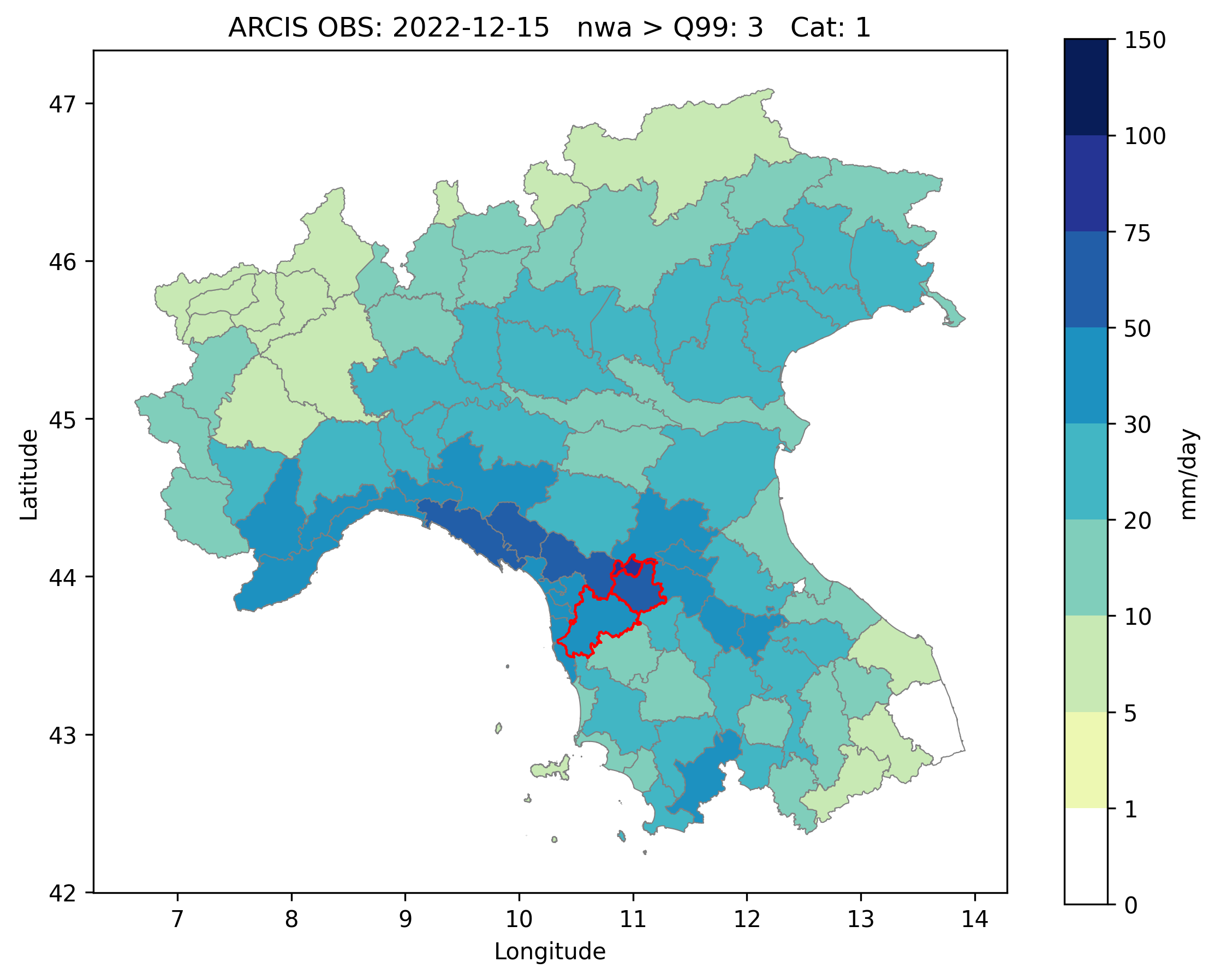}
\caption{24h observed precipitation during 15-12-2022 from the ARCIS dataset. Colour shading shows the precipitation accumulated between 00-24h spatially averaged over warning areas. In the three warning areas with read outlines the observed daily rainfall exceeded the 99th percentile of the daily accumulated precipitation}
\label{fig:obs_prec}
\end{figure}

Besides the verification statistic presented above, in this section we practically demonstrate the usage of MaLCoX, discussing one EPE event, which occurred after MalCoX's implementation at ARPAE in September 2022. The EPE occurred in December 2022 with three warning areas exceeding their respective 99th percentile of daily precipitation (Fig.\ref{fig:obs_prec}). Localized floodings were observed in Northern Tuscany due to continuous heavy stratiform rain, enhanced by orographic uplift, consistent with Cat1 classification. The city of Pistoia was particularly hit with 110mm of rain falling in 18 hours. Amongst the ranking of past EPEs, this can be considered moderate intensity, and small area (see Fig.\ref{fig:Vols} for comparison with other events). The ground effects were limited due to prevailing antecedent drought conditions, which resulted in good drainage in the mountain basins. Aside from the severity of the ground effects, this is an interesting exemplary case to illustrate how the forecasting system works even with relatively small amplitude events and to introduce some new forecast tools created to display the information content available with MaLCoX.

\begin{figure}[!ht]
\centering
\includegraphics[width=14cm]{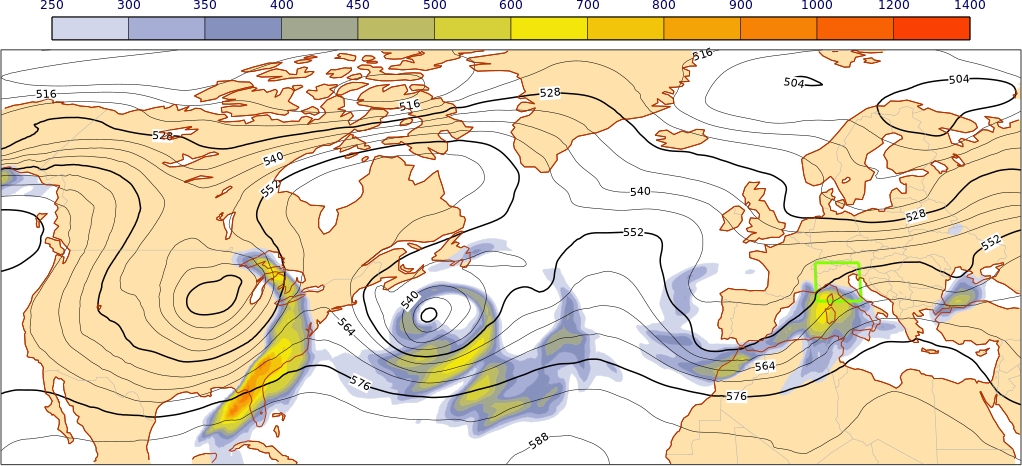}
\caption{Synoptic situation 2022-12-15 12:00 UTC. An EPE event occurred inside the test region depicted with the green box with floods in northern Tuscany. The map displays the gepotential height at 500 hPa (solid lines) and the associated magnitude of the instantaneous vertically integrated water vapour flow shaded according to the scale above in $kg s^{-1} m^{-1}$.}
\label{fig:Z500_example_Dec2015}
\end{figure}

The synoptic situation associated with this case study is shown in Fig.\ref{fig:Z500_example_Dec2015}. The EPE was associated with an amplifying Rossby wave centred over the Iberian peninsula, moving east. Further upstream there are regions of enhanced water vapour downstream of two other troughs, flowing in a low latitude zonally elongated region of high geopotential gradient. This configuration favours continuous replenishment of water vapour from the Atlantic to the Mediterranean basin. 

\begin{figure}[!ht]
     \centering
     \begin{subfigure}[b]{0.46\textwidth}
         \centering
         \includegraphics[width=\textwidth]{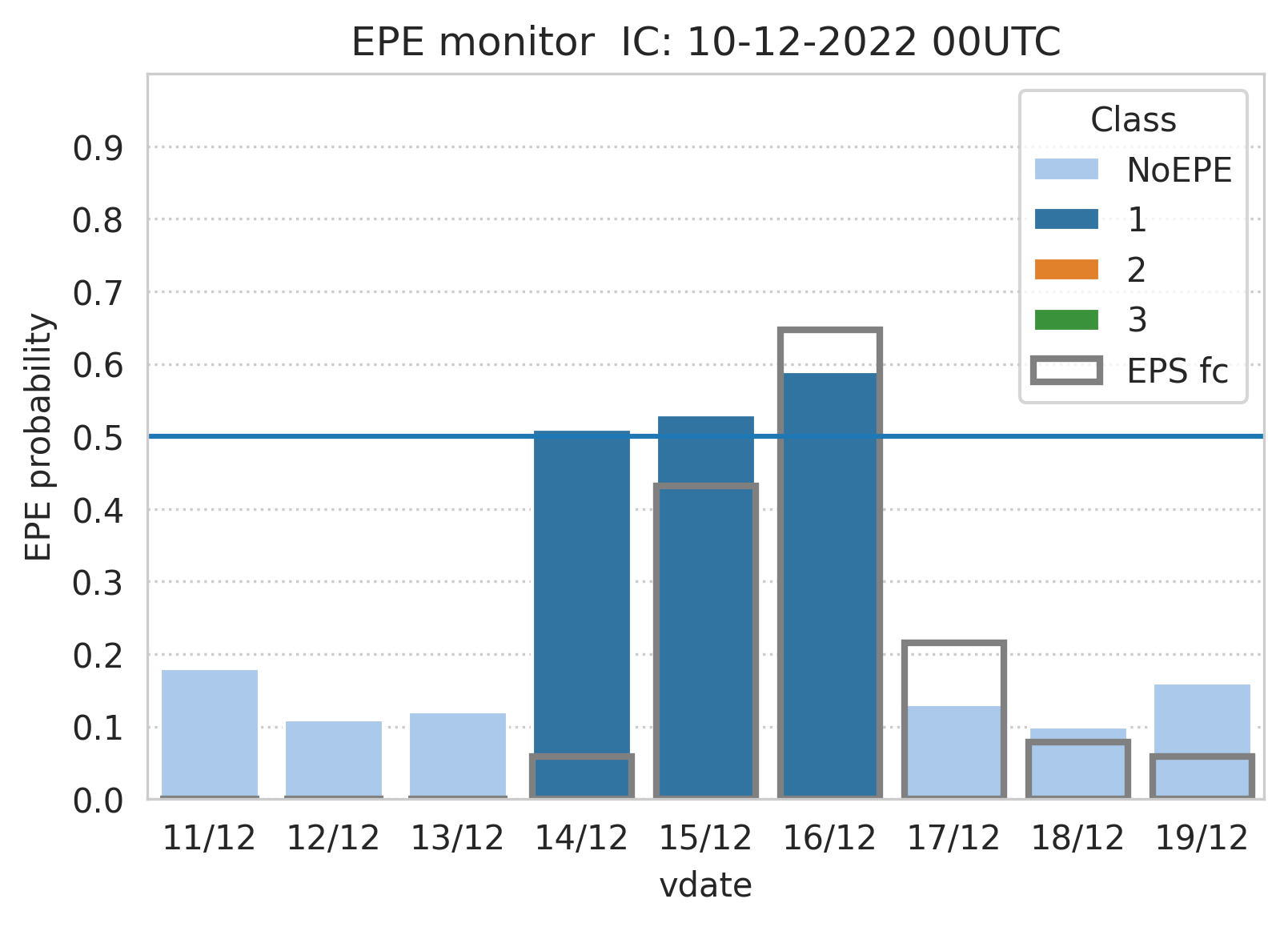}
         \caption{10 day forecast issued 2022-12-10}
         \label{fig:Monitor10}
     \end{subfigure}
     \hfill
     \begin{subfigure}[b]{0.46\textwidth}
         \centering
         \includegraphics[width=\textwidth]{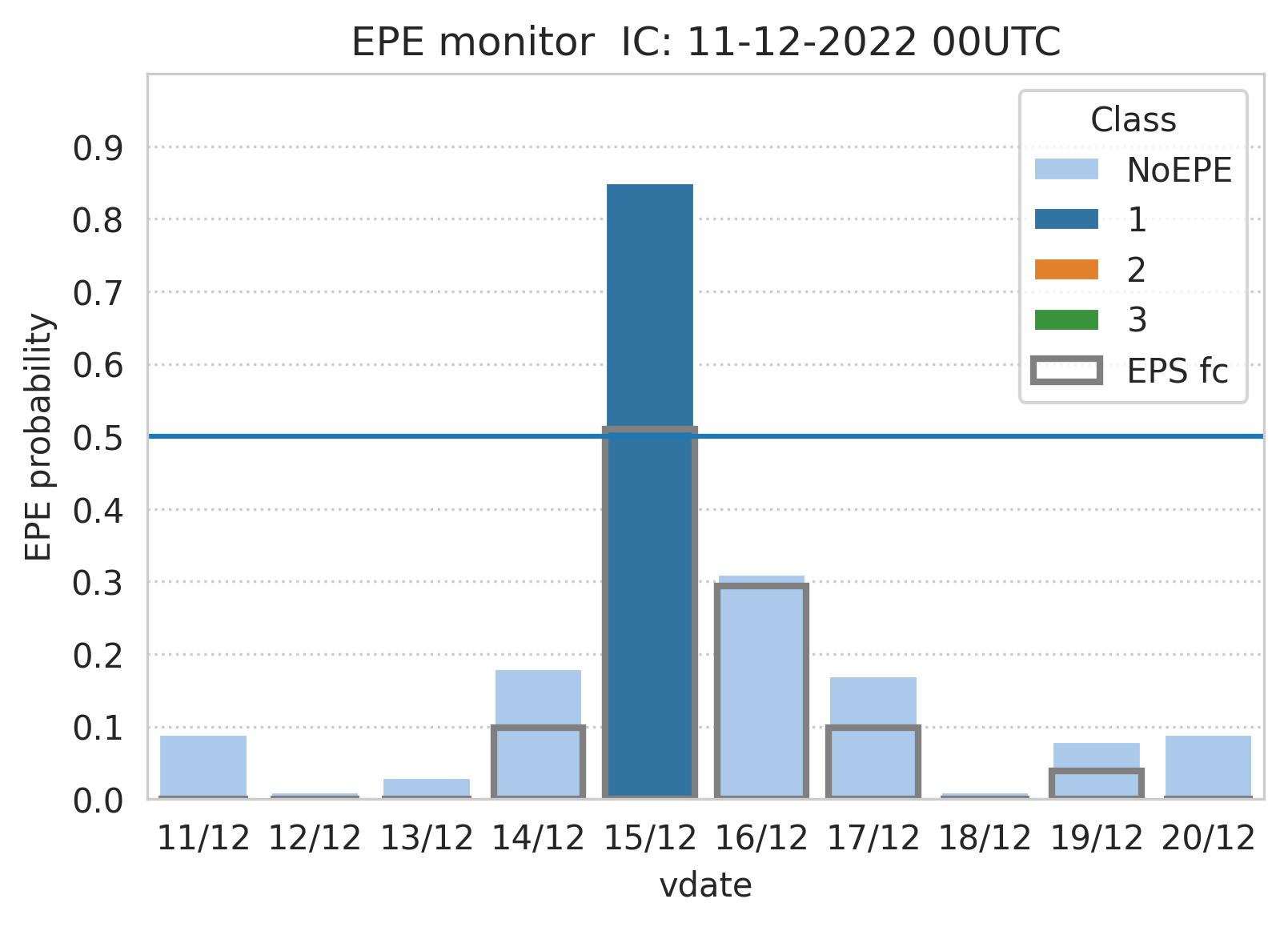}
         \caption{10 day forecast issued 2022-12-11}
         \label{fig:Monitor11}
     \end{subfigure}
     \caption{EPE monitor tool used to display MaLCoX predictions daily. The x-axes shows the forecast valid dates (day/month). The bars are showing the EPE probability computed by MaLCoX, coloured according to the predicted category (No EPE/cyan, Cat1/blue, Cat2/orange, Cat3/green). The probability obtained by processing the direct model output of the ENS members (empty grey bars) is also shown for comparison.}
        \label{fig:Monitors}
\end{figure}

\begin{figure}[!ht]
     \centering
     \begin{subfigure}[b]{0.46\textwidth}
         \centering
         \includegraphics[width=\textwidth]{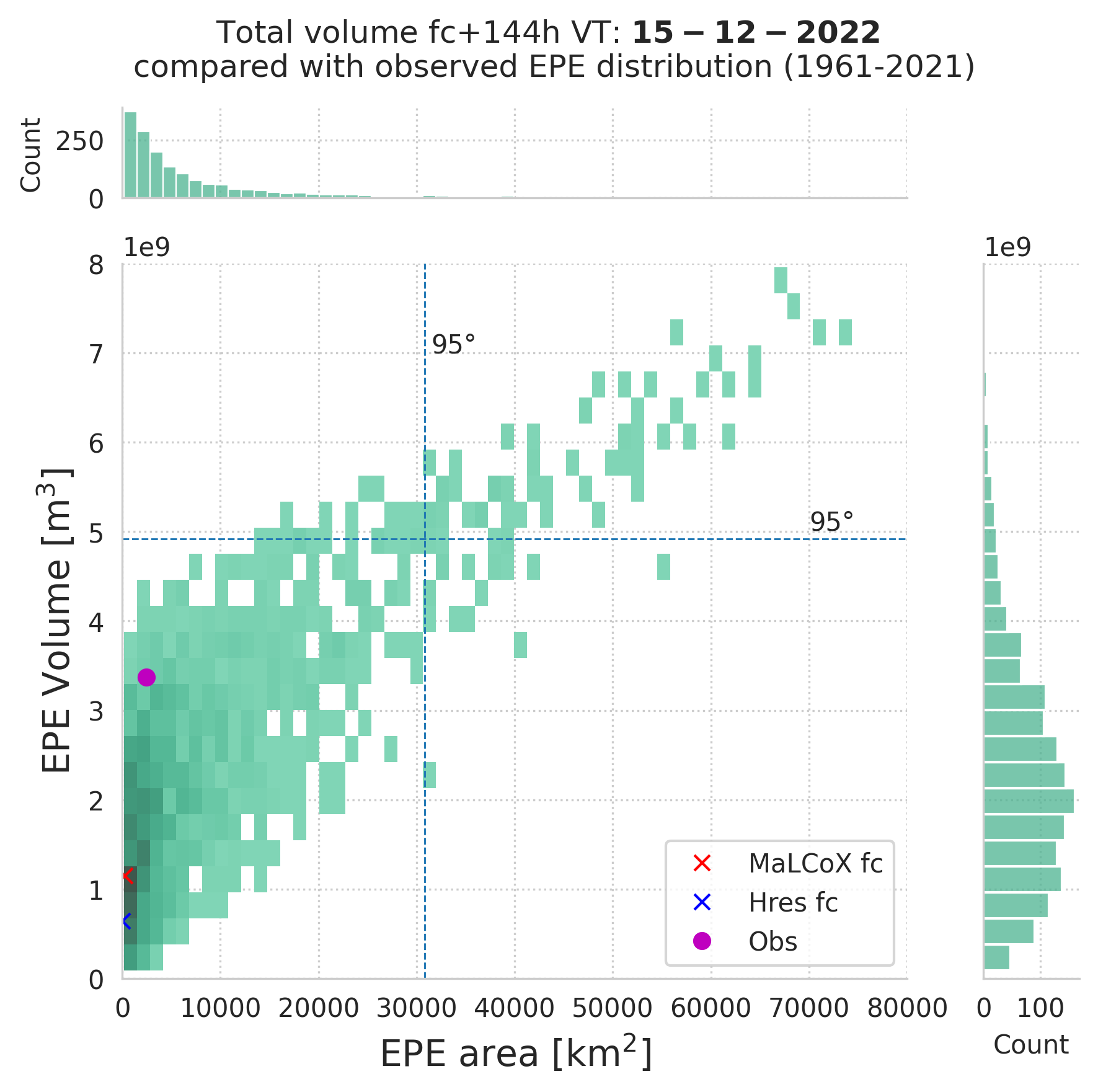}
         \caption{EPE rain volume prediction for 15-12-2022 issued 2022-12-10}
         \label{fig:Vol10}
     \end{subfigure}
     \hfill
     \begin{subfigure}[b]{0.46\textwidth}
         \centering
         \includegraphics[width=\textwidth]{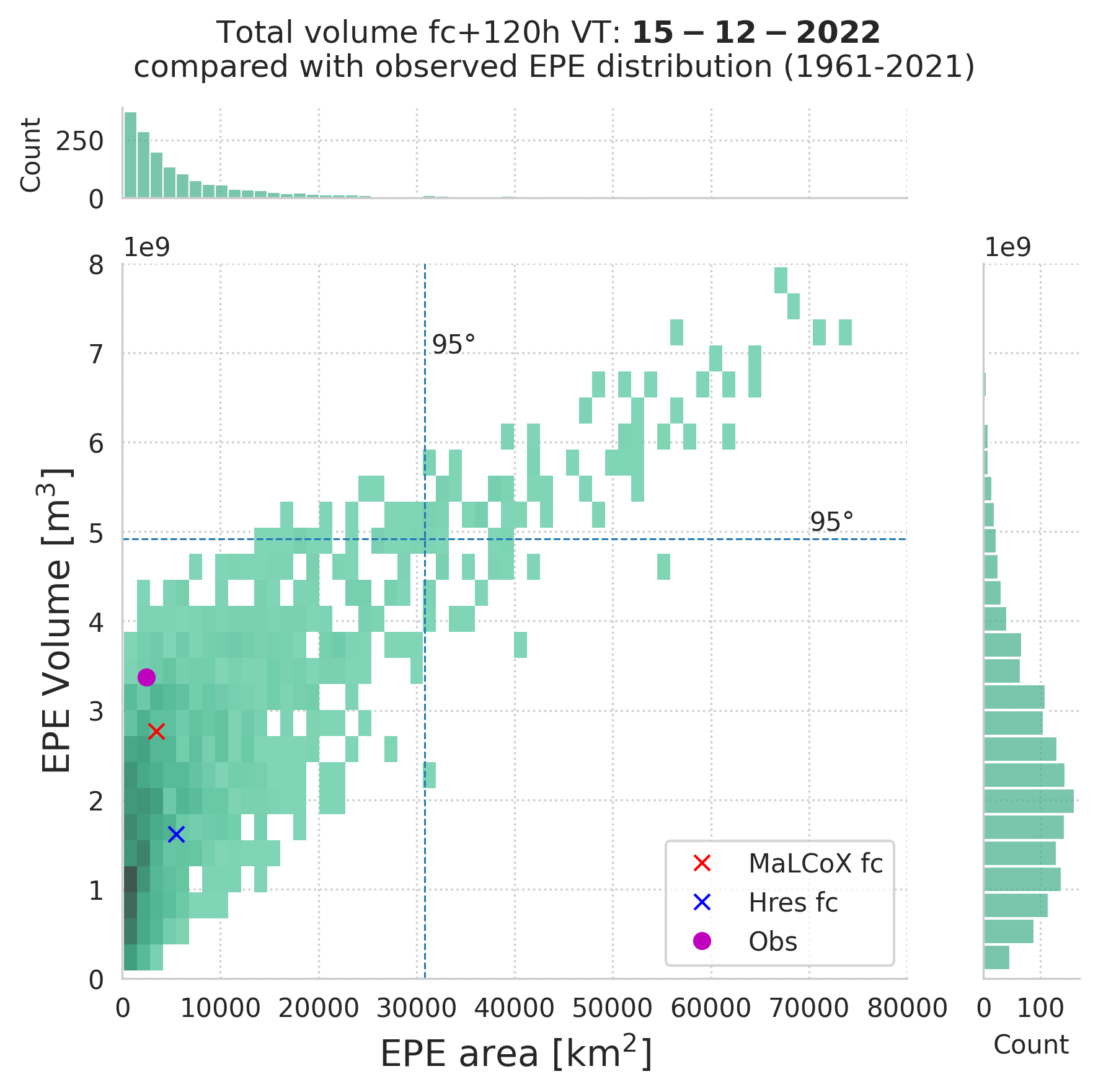}
         \caption{EPE rain volume prediction for 15-12-2022 issued 2022-12-11}
         \label{fig:Vol11}
     \end{subfigure}
     \caption{Total volume of rain and area extension of the EPE predicted by MaLCoX and IFS HRES. The shading of the cells represents the frequency distribution of Volume and Area of previous EPE observed between 1961 and 2022. Dashed lines are separating the plane in areas above or below the 95 percentile of the respective EPE distribution. For comparison, the violet dot shows the observed value}
        \label{fig:Vols}
\end{figure}

\begin{figure}[!ht]
     \centering
     \begin{subfigure}[b]{0.45\textwidth}
         \centering
         \includegraphics[width=\textwidth]{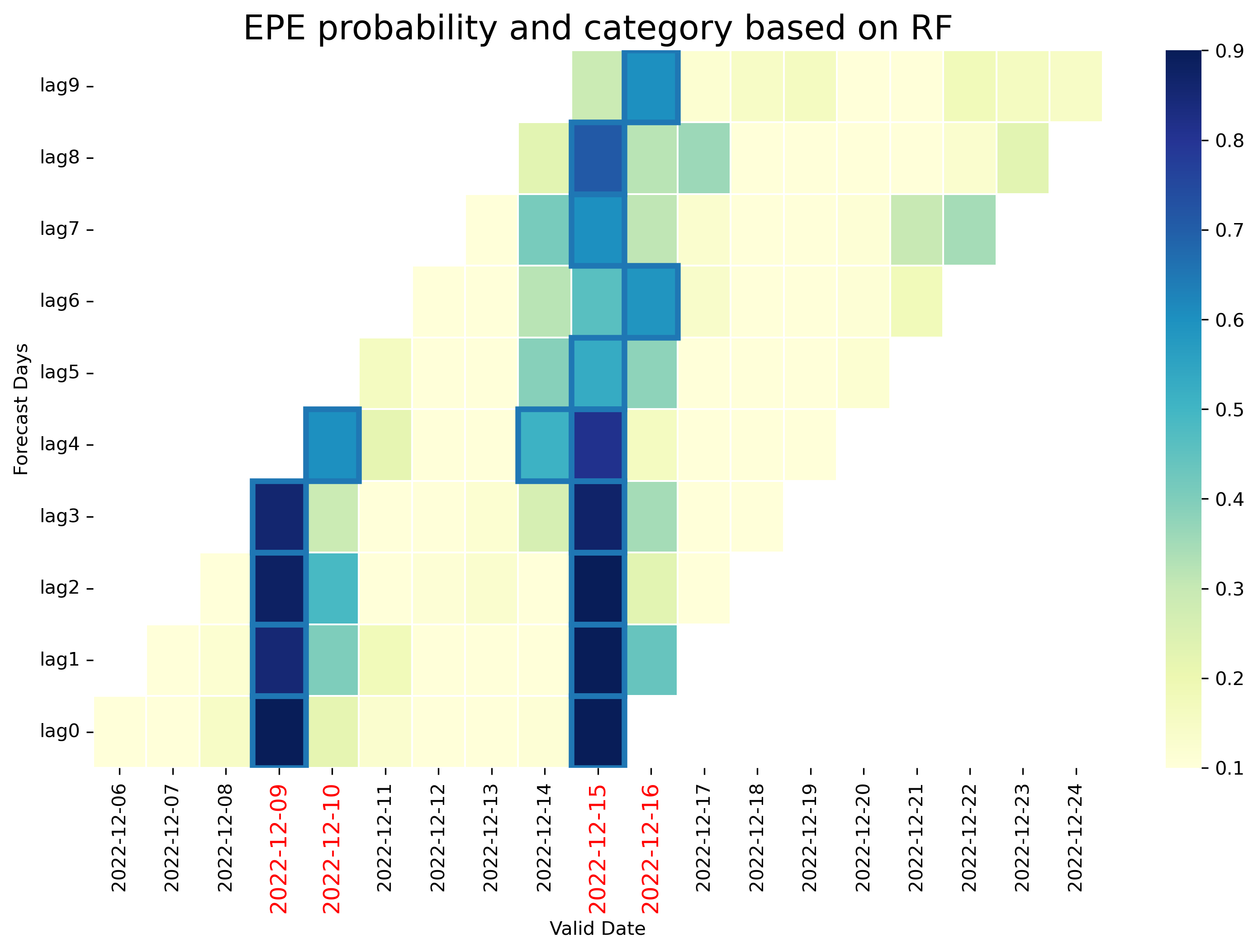}
         \caption{Probability and category indicated by the tick outline (Cat1/blue, Cat2/orange, Cat3/green). Dates marked in red indicate days with observed EPEs}
         \label{fig:RF_prob}
     \end{subfigure}
     \hfill
     \begin{subfigure}[b]{0.45\textwidth}
         \centering
         \includegraphics[width=\textwidth]{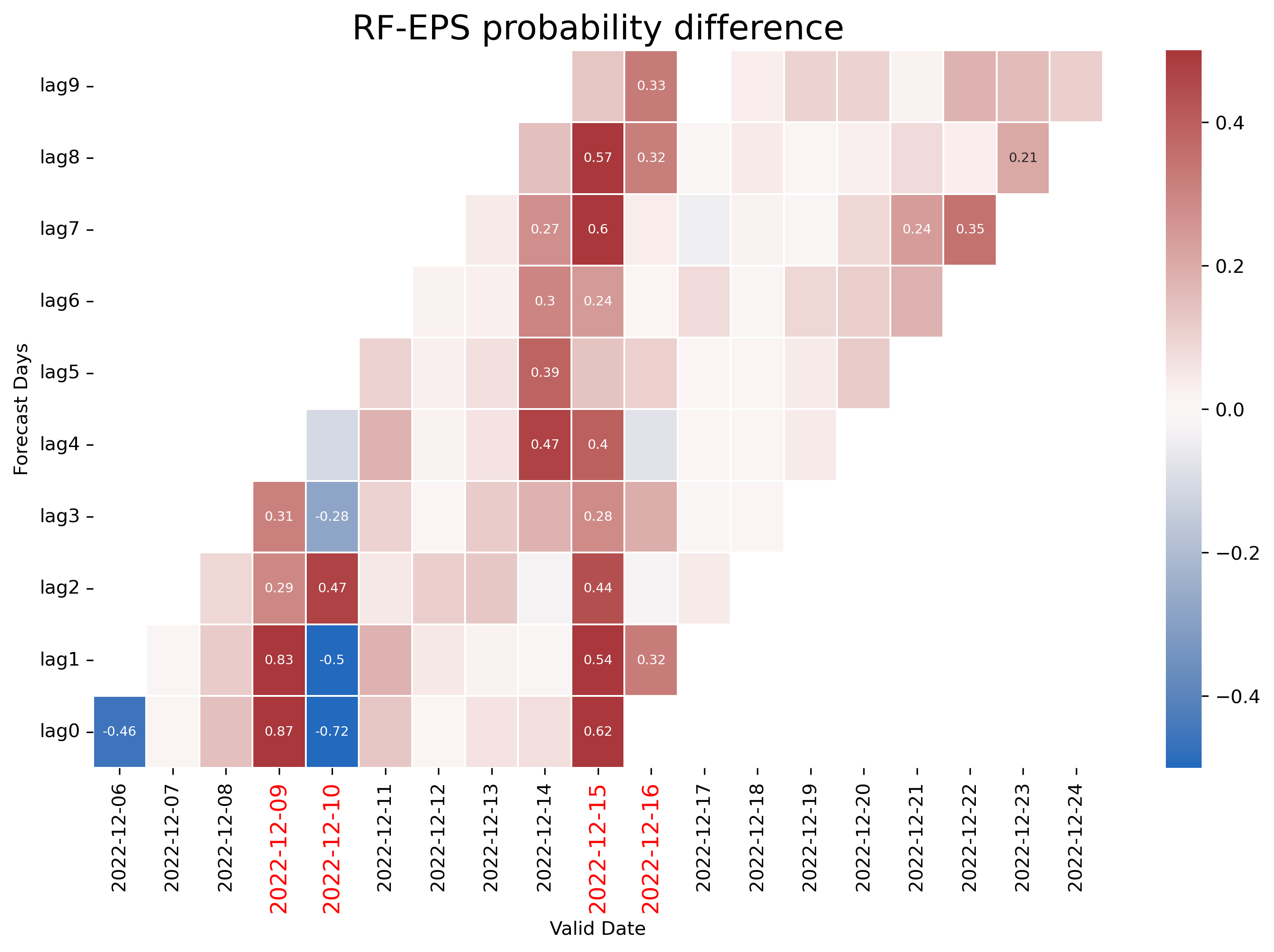}
         \caption{Difference in probability MaLCoX minus ENS according to legend. In boxes without annotation the difference is smaller in absolute terms than 0.2}
         \label{fig:Prob_difference}
     \end{subfigure}
     \caption{Probability heatmap for 20 days covering the case study}
        \label{fig:Prob_chess}
\end{figure}

In Fig.\ref{fig:Monitors} we show the EPE monitor display of two subsequent MaLCoX forecasts. This is our first alarm bell, showing the probabilistic output for an EPE in a ten-day forecast horizon. MaLCoX probability is displayed and compared against the corresponding probability obtained by ENS precipitation direct model output; the colour of the bars corresponds to the EPE category. In Fig.\ref{fig:Vols} we show a display referring to the expected intensity of the event in comparison with all EPEs which occurred before in terms of area above the 99 percentile and total volume of rain. This type of Volume-Area plot is automatically generated only if MaLCoX or ENS predict an EPE. As shown by Fig.\ref{fig:Monitors} and \ref{fig:Vols}, MaLCoX predicted the likelihood of an EPE Cat1 between 14 and 16 December many days in advance, but only from the forecast starting on 11/12, it started to point to the 15th of December as the day with the highest rainfall. MaLCox forecast issued on 11/12 showed a much higher and closer rain volume estimation than the one predicted by ECMWF raw output. Besides these quantitative measures, forecast consistency is another very desirable property of a reliable forecast system, particularly to increase trust. A further assessment of consistency can be obtained with a heatmap view of EPE probability. In Fig.\ref{fig:RF_prob}, we show the predicted probability and the EPE category by MaLCoX for a running window of 20 validating dates (x-axes) and forecast at different lead times (y-axes). MaLCoX became certain of the 15th from D+4 forecast (lag3) while at longer lead times the forecast was showing some inconsistency alternating days between 14 and 16 as possible EPEs. In Fig.\ref{fig:Prob_difference} we show a systematic difference of EPE probability predicted by MaLCoX minus the probabilities predicted by the ENS. MaLCoX correctly raised the probabilities (red boxes) compared to the ENS on the days with observed EPEs, except for a short-range forecast valid on the 10th of December. A general increase prevailed on the other days, but non significant, with differences smaller than 20\% (boxes without annotation). 

\begin{figure}[!ht]
     \centering
     \begin{subfigure}[b]{0.45\textwidth}
         \centering
         \includegraphics[width=\textwidth]{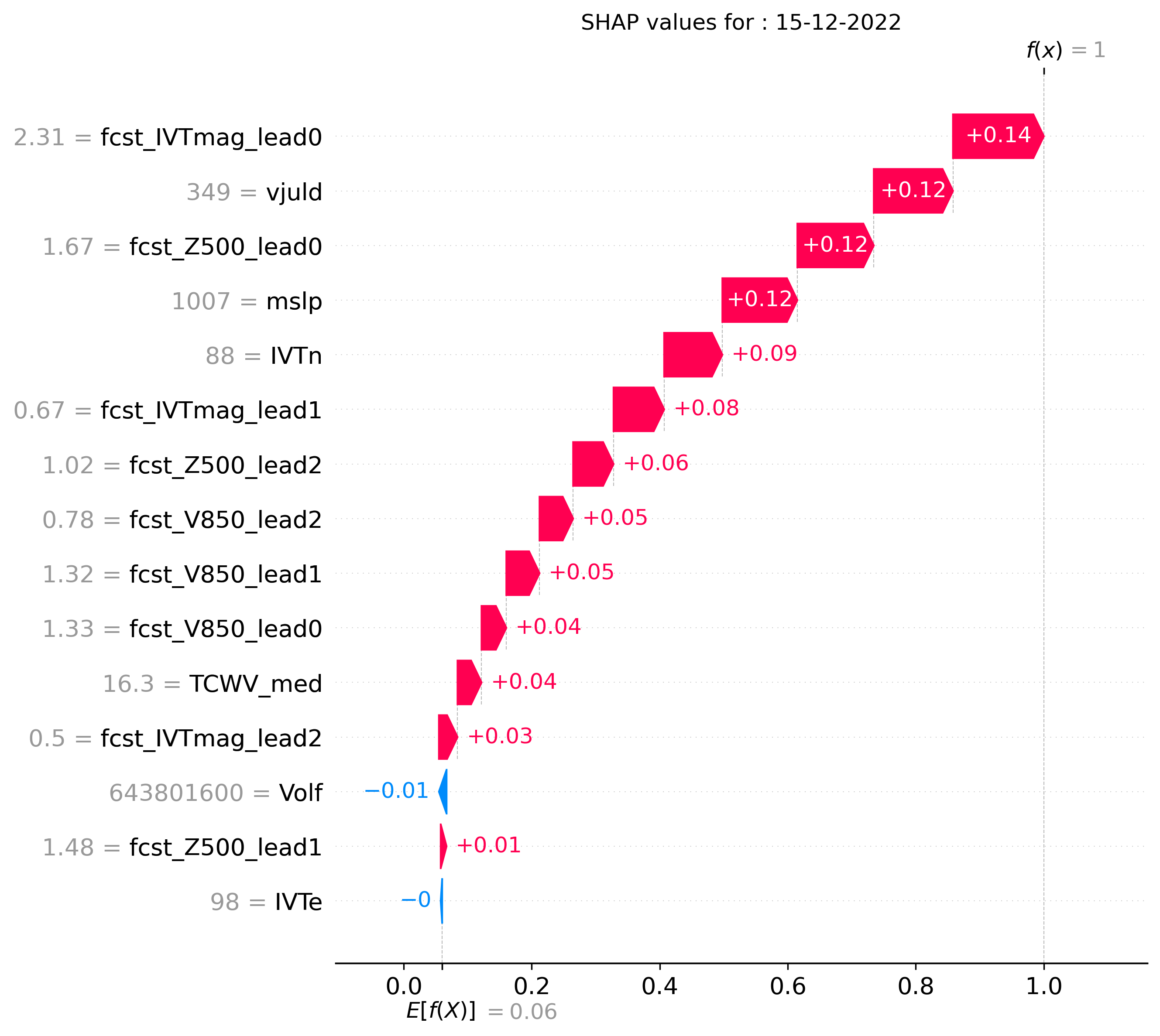}
         \caption{D+5 forecast initiated on 2022-12-10}
         \label{fig:Lag6_shap}
     \end{subfigure}
     \hfill
     \begin{subfigure}[b]{0.45\textwidth}
         \centering
         \includegraphics[width=\textwidth]{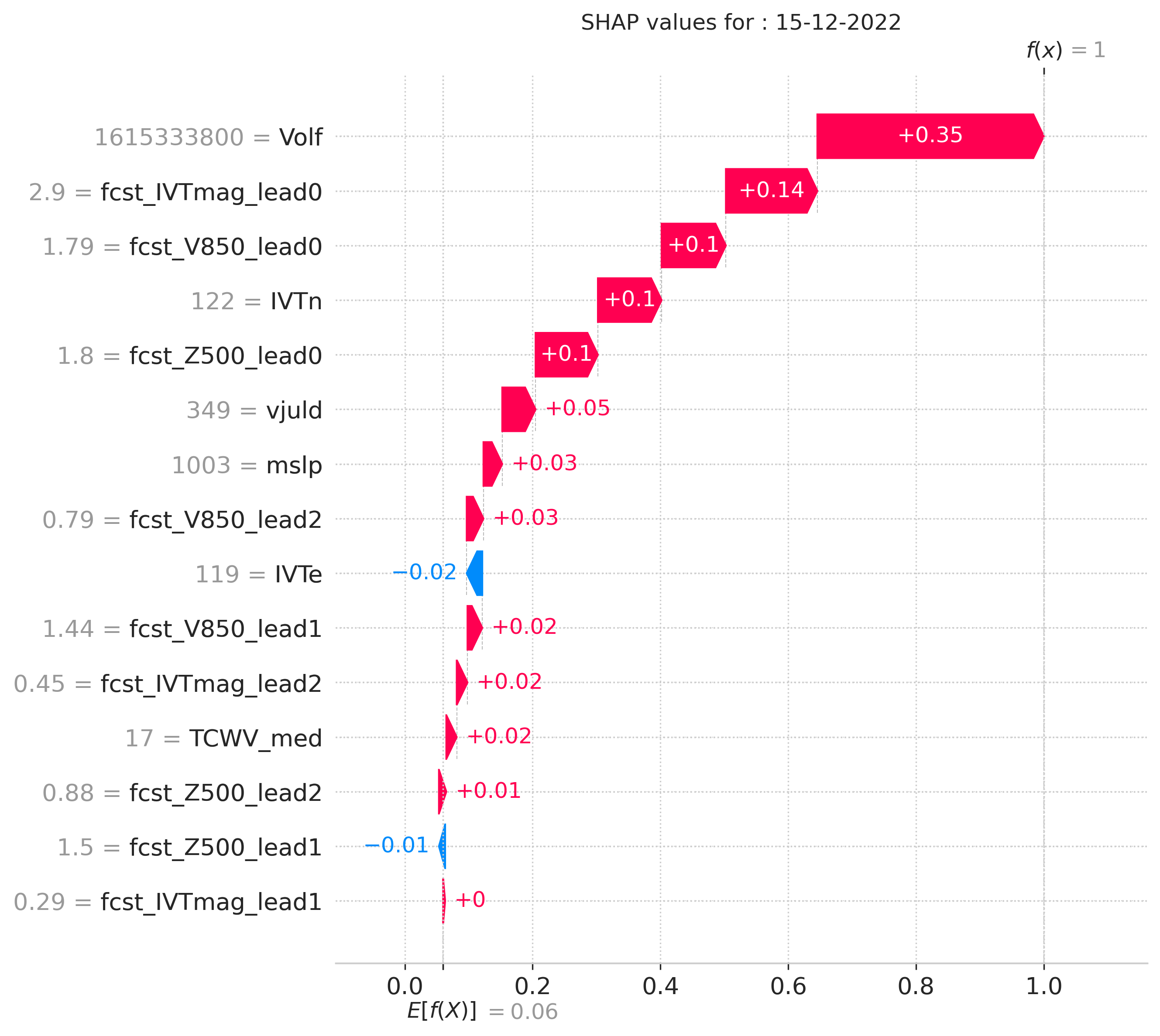}
         \caption{D+4 forecast initiated on 2022-12-11}
         \label{fig:Lag5_shap}
     \end{subfigure}
\caption{SHAP waterfall illustrating the additive contribution of each feature for two consecutive MaLCoX forecasts both validating on the 15th of December. In both forecasts, MaLCoX predicted an EPE but the result was driven by different predictors. The bottom of a waterfall plot starts as the climatological expected value of the model output for EPE = yes, and then each row shows how the positive (red) or negative (blue) contribution of each feature moves the value from the background value to the model output for this prediction. On the y-axes predictors' names are ranked according to their influence and their value is shown with grey numbers}
\label{fig:Shap}
\end{figure}

To explain the behaviour of MaLCoX and to convey the physical process which made the precipitation extreme in the forecast we adopted the Shapely additive explanation package \citep{Lundbergetal.2017}, already used for a similar application related to atmospheric composition forecast \citep{VegaGarcia2020ShapleyForecasting}. The waterfall representation, displayed in Fig.\ref{fig:Shap} shows the contribution of each feature in the two successive forecasts both validating on the 15th of December 2022. Starting from the background expectation E[f(X)], which is defined by the average frequency of the EPE in the forecast, 6\% at this lead time, the waterfall shows the incremental increase or decrease of the model output adding features until it reaches the outcome [f(X)=1] (EPE yes). In both forecasts, MaLCoX predicted an EPE but the decision was led by different predictors. In the forecast initiated on 10/12 (D+5) (Fig.\ref{fig:Lag6_shap}) the most important predictor is the IVTmag non-local index (lead time 0) followed by the day of the year (Vjuld) and Z500 non-local index (lead time 0), while in the forecast of next day (11/12) (Fig.\ref{fig:Lag5_shap}) the total rain volume predicted by HRES become the most important feature, followed by IVTmag non-local index (lead time 0) and V850 non-local index (lead time 0). In Fig.\ref{fig:Lag6_shap} it is interesting to note the slightly negative contribution of Volf due to HRES predicting a smaller rain amount compared with a typical EPE. Despite the lack of sufficient explicit rain, MaLCoX was still leading to an EPE due to the contribution of the large-scale component namely the non-local class. From D+4 and closer to the event, the ranking of predictors changes and Volf becomes the driving source of information with non-local predictors progressively going down in the ranking as the lead time reduces. Despite the predictors ranking changing case by case, this example illustrates well the typical complementary value of local and non-local predictors acting when the predictability of direct precipitation output is low. This improves the forecast skill in the medium range and the consistency preventing sudden jumps in the forecast. 

\section{Conclusion}
In this paper, we present a hybrid dynamical-statistical model (MaLCoX) which uses sequential random-forest models to combine different classes of forecast atmospheric predictors and improve predictions of extreme precipitation events (EPEs). These types of hybrid models represent an emerging class in the spectrum of weather forecasting systems. While previous studies have already reported hybrid predictions of short-term precipitation forecast, to our knowledge, no analogous examples of medium-range forecasting of extreme events exist, which is the target of our work. In the medium-range, non-local predictors -- scalar normalized indices representing large-scale anomalies preceding EPEs in the Euro-Atlantic sector -- are an important source of skill and a novel innovation of our approach. MaLCoX, trained with a low-resolution dataset with a 20 years equivalent period of 10-day forecast obtained from the ECMWF re-forecast dataset, shows a better performance than high-resolution ECMWF operational forecast in the prediction of EPEs over northern-central Italy, with a gain of about three days of forecast skill in the medium-range. A third innovative contribution is its interpretability: explaining the reason for a predicted outcome in terms of the relative contribution of different predictors. Understanding why a model makes a certain prediction can be as crucial as the prediction’s accuracy, especially when forecasters have to face rare extreme conditions and they need to gain trust in model output. Combining the results obtained by feature importance, scores on training and test datasets and results of feature attribution on single case studies, we show that predictors act in complementary ways throughout the forecasting period. At short forecast lead times, the hybrid model is mostly informed by the explicit precipitation field (total volume of rain) and local synoptic features, like IVTn. In the medium-range, and especially after lead time +144 hours, the cumulative effects of non-local predictors become predominant together with local synoptic features. The four-day forecast horizon is on average the crossing point where the redistribution of weights amongst the different feature components occurs. This information also has implications on predictability suggesting that, for the medium-range and even longer ranges, flow-patterns characteristics remain predictable and the information that can be extracted is useful to infer the likelihood of an EPE. Finally, we want to stress the positive role played by the non-local predictors class, here newly tested, in reducing the loss of predictability at longer time ranges. The results shown here refer to the first operational version that will be updated with further refinements. In particular, in the future upgrade of MaLCoX we plan to increase tenfold the training period using all ensemble members of the ENS re-forecast and to expand the prediction to all individual warning areas.

\section*{scripts and datasets}
The table containing observed aggregated daily precipitation over Northern and Central Italy warning areas (1961-2022) and days with EPE (yes/no) is available on the open data repository of ARPAE Emilia-Romagna: https://dati.datamb.it/dataset/serie-giornaliera-di-eventi-estremi-di-precipitazione-sul-centro-nord-italia



\printendnotes
\bibliography{references_Tex}

\begin{thebibliography}{31}
\expandafter\ifx\csname natexlab\endcsname\relax\def\natexlab#1{#1}\fi
\expandafter\ifx\csname url\endcsname\relax
  \def\url#1{\texttt{#1}}\fi
\expandafter\ifx\csname urlprefix\endcsname\relax\def\urlprefix{URL: }\fi

\bibitem[{Breiman(2001)}]{Breiman2001RandomForests}
Breiman, L. (2001) {Random forests}.
\newblock \textit{Machine Learning}, \textbf{45}, 5--32.

\bibitem[{Chapman et~al.(2022)Chapman, Monache, Alessandrini, Subramanian,
  Martin~Ralph, Xie, Lerch and Hayatbini}]{Chapman2022ProbabilisticLearning}
Chapman, W.~E., Monache, L.~D., Alessandrini, S., Subramanian, A.~C.,
  Martin~Ralph, F., Xie, S.~P., Lerch, S. and Hayatbini, N. (2022)
  {Probabilistic Predictions from Deterministic Atmospheric River Forecasts
  with Deep Learning}.
\newblock \textit{Monthly Weather Review}, \textbf{150}, 215--234.
\newblock
  \urlprefix\url{https://journals.ametsoc.org/view/journals/mwre/150/1/MWR-D-21-0106.1.xml}.

\bibitem[{Davis and Goadrich(2006)}]{Davis2006TheCurves}
Davis, J. and Goadrich, M. (2006) {The Relationship Between Precision-Recall
  and ROC Curves}.
\newblock \textit{Proceedings of the 23rd International Conference on
  MachineLearning, Pittsburgh,PA}.

\bibitem[{Dorrington et~al.(2023)Dorrington, Grams, Grazzini, Magnusson and
  Vitart}]{Dorrington2023Domino:Rainfall}
Dorrington, J., Grams, C., Grazzini, F., Magnusson, L. and Vitart, F. (2023)
  {Domino: A new framework for the automated identification of weather event
  precursors, demonstrated for European extreme rainfall}.
\newblock \textit{Quarterly Journal of the Royal Meteorological Society}.
\newblock
  \urlprefix\url{https://onlinelibrary.wiley.com/doi/full/10.1002/qj.4622
  https://onlinelibrary.wiley.com/doi/abs/10.1002/qj.4622
  https://rmets.onlinelibrary.wiley.com/doi/10.1002/qj.4622}.

\bibitem[{Duffourg and Ducrocq(2011)}]{Duffourg2011OriginFrance}
Duffourg, F. and Ducrocq, V. (2011) {Origin of the moisture feeding the heavy
  precipitating systems over southeastern France}.
\newblock \textit{Natural Hazards and Earth System Science}, \textbf{11},
  1163--1178.

\bibitem[{Espeholt et~al.(2022)Espeholt, Agrawal, S{\o}nderby, Kumar, Heek,
  Bromberg, Gazen, Carver, Andrychowicz, Hickey, Bell and
  Kalchbrenner}]{Espeholt2022DeepForecasts}
Espeholt, L., Agrawal, S., S{\o}nderby, C., Kumar, M., Heek, J., Bromberg, C.,
  Gazen, C., Carver, R., Andrychowicz, M., Hickey, J., Bell, A. and
  Kalchbrenner, N. (2022) {Deep learning for twelve hour precipitation
  forecasts}.
\newblock \textit{Nature Communications}.
\newblock \urlprefix\url{https://doi.org/10.1038/s41467-022-32483-x}.

\bibitem[{Frnda et~al.(2022)Frnda, Durica, Rozhon, Vojtekova, Nedoma and
  Martinek}]{Frnda2022ECMWFLearning}
Frnda, J., Durica, M., Rozhon, J., Vojtekova, M., Nedoma, J. and Martinek, R.
  (2022) {ECMWF short-term prediction accuracy improvement by deep learning}.
\newblock \textit{Scientific Reports}, \textbf{12}.
\newblock \urlprefix\url{https://doi.org/10.1038/s41598-022-11936-9}.

\bibitem[{Grazzini(2007)}]{Grazzini2007PredictabilityAlps}
Grazzini, F. (2007) {Predictability of a large-scale flow conducive to extreme
  precipitation over the western Alps}.
\newblock \textit{Meteorology and Atmospheric Physics}, \textbf{95}, 123--138.
\newblock \urlprefix\url{http://link.springer.com/10.1007/s00703-006-0205-8}.

\bibitem[{Grazzini(2021)}]{ediss28219}
--- (2021) {Extreme precipitation in Northern Italy}.
\newblock \urlprefix\url{http://nbn-resolving.de/urn:nbn:de:bvb:19-282191}.

\bibitem[{Grazzini et~al.(2020{\natexlab{a}})Grazzini, Craig, Keil, Antolini
  and Pavan}]{Grazzini2020ExtremeTechniques}
Grazzini, F., Craig, G.~C., Keil, C., Antolini, G. and Pavan, V.
  (2020{\natexlab{a}}) {Extreme precipitation events over northern Italy. Part
  I: A systematic classification with machine‐learning techniques}.
\newblock \textit{Quarterly Journal of the Royal Meteorological Society},
  \textbf{146}, 69--85.
\newblock \urlprefix\url{https://onlinelibrary.wiley.com/doi/10.1002/qj.3635}.

\bibitem[{Grazzini et~al.(2020{\natexlab{b}})Grazzini, Fragkoulidis, Pavan and
  Antolini}]{Grazzini2020TheItaly}
Grazzini, F., Fragkoulidis, G., Pavan, V. and Antolini, G. (2020{\natexlab{b}})
  {The 1994 Piedmont flood: an archetype of extreme precipitation events in
  Northern Italy}.
\newblock \textit{Bulletin of Atmospheric Science and Technology}, 1--13.
\newblock \urlprefix\url{https://doi.org/10.1007/s42865-020-00018-1}.

\bibitem[{Grazzini et~al.(2021)Grazzini, Fragkoulidis, Teubler, Wirth and
  Craig}]{Grazzini2021ExtremePrecursors}
Grazzini, F., Fragkoulidis, G., Teubler, F., Wirth, V. and Craig, G.~C. (2021)
  {Extreme precipitation events over northern Italy. Part II: Dynamical
  precursors}.
\newblock \textit{Quarterly Journal of the Royal Meteorological Society},
  qj.3969.
\newblock \urlprefix\url{https://onlinelibrary.wiley.com/doi/10.1002/qj.3969}.

\bibitem[{Grazzini and Vitart(2015)}]{Grazzini2015AtmosphericPackets}
Grazzini, F. and Vitart, F. (2015) {Atmospheric predictability and Rossby wave
  packets}.
\newblock \textit{Quarterly Journal of the Royal Meteorological Society},
  \textbf{141}.

\bibitem[{Hill et~al.(2020)Hill, Herman and
  Schumacher}]{Hill2020ForecastingForests}
Hill, A.~J., Herman, G.~R. and Schumacher, R.~S. (2020) {Forecasting Severe
  Weather with Random Forests}.
\newblock \textit{Monthly Weather Review}, \textbf{148}, 2135--2161.
\newblock
  \urlprefix\url{https://journals.ametsoc.org/view/journals/mwre/148/5/mwr-d-19-0344.1.xml}.

\bibitem[{Khodayar et~al.(2021)Khodayar, Davolio, Di~Girolamo,
  Lebeaupin~Brossier, Flaounas, Fourrie, Lee, Ricard, Vie, Bouttier,
  Caldas-Alvarez and Ducrocq}]{Khodayar2021OverviewHyMeX}
Khodayar, S., Davolio, S., Di~Girolamo, P., Lebeaupin~Brossier, C., Flaounas,
  E., Fourrie, N., Lee, K.~O., Ricard, D., Vie, B., Bouttier, F.,
  Caldas-Alvarez, A. and Ducrocq, V. (2021) {Overview towards improved
  understanding of the mechanisms leading to heavy precipitation in the western
  Mediterranean: Lessons learned from HyMeX}.
\newblock \textit{Atmospheric Chemistry and Physics}, \textbf{21},
  17051--17078.

\bibitem[{Khodayar et~al.(2022)Khodayar, Pastor, Valiente, Benet{\'{o}} and
  Ehmele}]{Khodayar2022WhatMediterranean}
Khodayar, S., Pastor, F., Valiente, J.~A., Benet{\'{o}}, P. and Ehmele, F.
  (2022) {What causes a heavy precipitation period to become extreme? The
  exceptional October of 2018 in the Western Mediterranean}.
\newblock \textit{Weather and Climate Extremes}, \textbf{38}, 100493.

\bibitem[{Lundberg et~al.(2017)Lundberg, Allen and Lee}]{Lundbergetal.2017}
Lundberg, S.~M., Allen, P.~G. and Lee, S.-I. (2017) {A Unified Approach to
  Interpreting Model Predictions}.
\newblock \textit{Tech. rep.}
\newblock \urlprefix\url{https://github.com/slundberg/shap}.

\bibitem[{Magnusson et~al.(2021)Magnusson, Hewson and
  Lavers}]{Magnusson2021WindstormEurope}
Magnusson, L., Hewson, T. and Lavers, D. (2021) {Windstorm Alex affected large
  parts of Europe}.
\newblock \textit{ECMWF Newsletter N. 166}, 4--5.

\bibitem[{Martius et~al.(2008)Martius, Schwierz and
  Davies}]{Martius2008Far-upstreamSouth-side}
Martius, O., Schwierz, C. and Davies, H.~C. (2008) {Far-upstream precursors of
  heavy precipitation events on the Alpine south-side}.
\newblock \textit{Quarterly Journal of the Royal Meteorological Society},
  \textbf{134}, 417--428.
\newblock \urlprefix\url{http://doi.wiley.com/10.1002/qj.229}.

\bibitem[{Mastrantonas et~al.(2022)Mastrantonas, Magnusson, Pappenberger and
  Matschullat}]{Mastrantonas2022WhatForecasts}
Mastrantonas, N., Magnusson, L., Pappenberger, F. and Matschullat, J. (2022)
  {What do large-scale patterns teach us about extreme precipitation over the
  Mediterranean at medium-and extended-range forecasts?}
\newblock
  \urlprefix\url{https://rmets.onlinelibrary.wiley.com/doi/10.1002/qj.4236}.

\bibitem[{Pavan et~al.(2019)Pavan, Antolini, Barbiero, Berni, Brunier,
  Cacciamani, Cagnati, Cazzuli, Cicogna, De~Luigi, Di~Carlo, Francioni,
  Maraldo, Marigo, Micheletti, Onorato, Panettieri, Pellegrini, Pelosini,
  Piccinini, Ratto, Ronchi, Rusca, Sofia, Stelluti, Tomozeiu and
  Torrigiani~Malaspina}]{Pavan2019High19612015}
Pavan, V., Antolini, G., Barbiero, R., Berni, N., Brunier, F., Cacciamani, C.,
  Cagnati, A., Cazzuli, O., Cicogna, A., De~Luigi, C., Di~Carlo, E., Francioni,
  M., Maraldo, L., Marigo, G., Micheletti, S., Onorato, L., Panettieri, E.,
  Pellegrini, U., Pelosini, R., Piccinini, D., Ratto, S., Ronchi, C., Rusca,
  L., Sofia, S., Stelluti, M., Tomozeiu, R. and Torrigiani~Malaspina, T. (2019)
  {High resolution climate precipitation analysis for north-central Italy,
  1961–2015}.
\newblock \textit{Climate Dynamics}, \textbf{52}, 3435--3453.

\bibitem[{Pedregosa et~al.(2011)Pedregosa, Varoquaux, Gramfort, Michel,
  Thirion, Grisel, Blondel, Prettenhofer, Weiss, Dubourg, Vanderplas, Passos,
  Cournapeau, Brucher, Perrot and Duchesnay}]{Pedregosa2011Scikit-learn:Python}
Pedregosa, F., Varoquaux, G., Gramfort, A., Michel, V., Thirion, B., Grisel,
  O., Blondel, M., Prettenhofer, P., Weiss, R., Dubourg, V., Vanderplas, J.,
  Passos, A., Cournapeau, D., Brucher, M., Perrot, M. and Duchesnay, E. (2011)
  {Scikit-learn: Machine Learning in Python}.
\newblock \textit{Journal of Machine Learning Research}, \textbf{12},
  2825--2830.
\newblock \urlprefix\url{http://jmlr.org/papers/v12/pedregosa11a.html}.

\bibitem[{Rudari et~al.(2005)Rudari, Entekhabi and
  Roth}]{Rudari2005Large-scaleItaly}
Rudari, R., Entekhabi, D. and Roth, G. (2005) {Large-scale atmospheric patterns
  associated with mesoscale features leading to extreme precipitation events in
  Northwestern Italy}.
\newblock \textit{Advances in Water Resources}, \textbf{28}, 601--614.

\bibitem[{Seneviratne and {et al.}(2021)}]{Seneviratne2021WeatherChange}
Seneviratne, S. and {et al.} (2021) {Weather and Climate Extreme Events in a
  Changing Climate. In Climate Change 2021: The Physical Science Basis.
  Contribution of Working Group I to the Sixth Assessment Report of the
  Intergovernmental Panel on Climate Change}.
\newblock \textit{Cambridge University Press}, 1513--1766.

\bibitem[{Sioni et~al.(2023)Sioni, Davolio, Grazzini and
  Giovannini}]{Sioni2023RevisitingItaly}
Sioni, F., Davolio, S., Grazzini, F. and Giovannini, L. (2023) {Revisiting the
  atmospheric dynamics of the two century floods over north-eastern Italy}.
\newblock \textit{Atmospheric Research}, \textbf{286}, 106662.
\newblock
  \urlprefix\url{https://linkinghub.elsevier.com/retrieve/pii/S0169809523000595}.

\bibitem[{de~Sousa~Ara{\'{u}}jo et~al.(2022)de~Sousa~Ara{\'{u}}jo, Silva and
  Z{\'{a}}rate}]{deSousaAraujo2022ExtremeBrazil}
de~Sousa~Ara{\'{u}}jo, A., Silva, A.~R. and Z{\'{a}}rate, L.~E. (2022) {Extreme
  precipitation prediction based on neural network model – A case study for
  southeastern Brazil}.
\newblock \textit{Journal of Hydrology}, \textbf{606}, 127454.

\bibitem[{Tramblay and Somot(2018)}]{Tramblay2018FutureMediterranean}
Tramblay, Y. and Somot, S. (2018) {Future evolution of extreme precipitation in
  the Mediterranean}.
\newblock \textit{Climatic Change}, \textbf{151}, 289--302.

\bibitem[{Tsonevsky(2015)}]{Tsonevsky2015NewConvection.}
Tsonevsky, I. (2015) {New EFI parameters for forecasting severe convection. }.
\newblock \textit{ECMWF Newsletter N.144}, 27--32.

\bibitem[{Vega~Garc{\'{i}}a and
  Aznarte(2020)}]{VegaGarcia2020ShapleyForecasting}
Vega~Garc{\'{i}}a, M. and Aznarte, J.~L. (2020) {Shapley additive explanations
  for NO2 forecasting}.
\newblock \textit{Ecological Informatics}, \textbf{56}, 101039.

\bibitem[{Vitart et~al.(2019)Vitart, Balsamo, Bidlot, Lang, Tsonevsky,
  Richardson and Alonso-Balmaseda}]{Vitart2019}
Vitart, F., Balsamo, G., Bidlot, J.-R., Lang, S., Tsonevsky, I., Richardson, D.
  and Alonso-Balmaseda, M. (2019) {Use of ERA5 to Initialize Ensemble
  Re-forecasts}.
\newblock \urlprefix\url{https://www.ecmwf.int/node/18872}.

\bibitem[{Whan and Schmeits(2018)}]{Whan2018ComparingMethods}
Whan, K. and Schmeits, M. (2018) {Comparing Area Probability Forecasts of
  (Extreme) Local Precipitation Using Parametric and Machine Learning
  Statistical Postprocessing Methods}.
\newblock \textit{Monthly Weather Review}, \textbf{146}, 3651--3673.
\newblock
  \urlprefix\url{https://journals.ametsoc.org/view/journals/mwre/146/11/mwr-d-17-0290.1.xml}.

\end{thebibliography}

\begin{biography}[example-image-1x1]{A.~One}
Please check with the journal's author guidelines whether author biographies are required. They are usually only included for review-type articles, and typically require photos and brief biographies (up to 75 words) for each author.
\bigskip
\bigskip
\end{biography}

\graphicalabstract{figures/stat/Schema_MaLCoX.png}{Improving forecast of precipitation extremes over Northern and Central Italy using machine learning. F.Grazzini*, J.Dorrington, C.Grams, G.Craig, L.Magnusson, F.Vitart.\\
In this study, we show that by blending direct model outputs with past event statistics, local and non-local predictors, through the usage of a machine learning hybrid models, we obtain more reliable predictions at longer ranges. For each event, we also gain knowledge of the physical processes making it an extreme}

\end{document}